\documentclass[sigconf, nonacm]{acmart}

\renewcommand\footnotetextcopyrightpermission[1]{} %
\usepackage[utf8]{inputenc}%

\usepackage{algorithmic}
\usepackage{graphicx}
\usepackage{pgf}
\usepackage{tikz}
\usetikzlibrary{arrows,patterns,decorations.pathmorphing,decorations.pathreplacing,backgrounds,positioning,fit,matrix}
\usetikzlibrary{shapes,calc}

\usepackage{dsfont}

\usepackage[ruled]{algorithm2e} %

\usepackage{xcolor}
\definecolor{sron0}{HTML}{332288}
\definecolor{sron1}{HTML}{88CCEE}
\definecolor{sron2}{HTML}{117733}
\definecolor{sron3}{HTML}{DDCC77}
\definecolor{sron4}{HTML}{CC6677}

\usepackage{hyperref}
\usepackage{textcomp}
\usepackage{csquotes}
\usepackage{xcolor}
\usepackage{xspace}
\newcommand*{\eg}{e.g.\@\xspace}
\newcommand*{\ie}{i.e.\@\xspace}

\newcommand*{\lngraphsymb}{\mathcal{G}}
\newcommand*{\lnpubgraphsymb}{\mathcal{G}_{pub}}
\newcommand*{\lnrbalsymb}{\mathcal{R}_\lnbalsymb}
\newcommand*{\lnrlocksymb}{\mathcal{R}_\mathbf{\Delta}}
\newcommand*{\lnrcapsymb}{\mathcal{R}_\lncapsymb}
\newcommand*{\lnverticessymb}{\mathcal{V}}
\newcommand*{\lnedgessymb}{\mathcal{E}}
\newcommand*{\lnbalsymb}{\mathbf{bal}}
\newcommand*{\lncapsymb}{\mathbf{cap}}
\newcommand*{\lnfeesymb}{\mathbf{fee}}
\newcommand*{\lnlocksymb}{\mathbf{\Delta_{tl}}}
\newcommand*{\lnmaxlocksymb}{\mathbf{\Delta_{max}}}
\newcommand*{\lnlatsymb}{\mathbf{lat}}
\newcommand*{\lnroutealgsymb}{\mathfrak{R}}
\newcommand*{\lnestimatorsymb}{\mathfrak{M}}
\newcommand*{\lndatasymb}{\mathcal{D}}

\newcommand{\lnbal}[3]{\lnbalsymb(#1,#2,#3)}
\newcommand{\lncap}[1]{\lncapsymb(#1)}
\newcommand{\lnfee}[4]{\lnfeesymb(#4|#1,#2,#3)}
\newcommand{\lnlock}[3]{\lnlocksymb(#1,#2,#3)}
\newcommand{\lnlat}[1]{\lnlatsymb(#1)}
\newcommand{\lnroutealg}[2]{\lnroutealgsymb(#2|#1)}
\newcommand{\lnestimator}[5]{\lnestimatorsymb_#4(#5|#1,#2,#3)}

\usepackage{tabularx}
\usepackage{booktabs}
\AtBeginDocument{%
  \providecommand\BibTeX{{%
    \normalfont B\kern-0.5em{\scshape i\kern-0.25em b}\kern-0.8em\TeX}}}

\setcopyright{none}

\begin{document}

\title[Timing Attacks on Privacy in Payment Channel Networks]{Counting Down
Thunder:\\Timing Attacks on Privacy in Payment Channel Networks}

\author{Elias Rohrer}
\email{elias.rohrer@tu-berlin.de}
\orcid{0000-0003-1651-9838}
\affiliation{%
  \institution{Technical University of Berlin}
  \city{Berlin, Germany}
}

\author{Florian Tschorsch}
\email{florian.tschorsch@tu-berlin.de}
\affiliation{%
  \institution{Technical University of Berlin}
  \city{Berlin, Germany}
}

\begin{abstract}
    The Lightning Network is a scaling solution for
    Bitcoin that promises to enable rapid and
    private payment processing.
    In Lightning, multi-hop payments are secured by utilizing Hashed Time-Locked Contracts
    (HTLCs) and encrypted on the network layer by an onion routing scheme to
    avoid information leakage to intermediate nodes. In this work, we however
    show that the privacy guarantees of the Lightning Network may be subverted
    by an on-path adversary conducting timing attacks on the HTLC
    state negotiation messages. To this end, we provide estimators that
    enable an adversary to reduce the anonymity set and infer the likeliest
    payment endpoints.
    We developed a proof-of-concept measurement node that shows the
    feasibility of attaining time differences and evaluate the
    adversarial success in model-based network simulations. We find that
    controlling a small number malicious nodes is sufficient to observe a large
    share of all payments, emphasizing the relevance of the on-path
    adversary model. Moreover, we
    show that adversaries of different magnitudes could employ timing-based
    attacks to deanonymize payment endpoints with high precision and recall.
\end{abstract}

\maketitle

\section{Introduction}\label{sec:intro}
In the years since its introduction, Bitcoin~\cite{nakamoto2008bitcoin} has
proven the feasibility of conducting financial transactions
without a centralized clearing house.
While it enjoys increasing
popularity around the world, it has been shown that its open and transparent design
severely limits the achievable transaction throughput~\cite{croman2016scalingblockchains}
as well as the privacy on, both,
the consensus~\cite{meiklejohn2013fistful} and networking layers~\cite{fanti2017deanon}.

Second-layer solutions, such as Bitcoin's Lightning Network~\cite{poon2015bitcoin},
promise to mitigate these shortcomings by establishing a network of off-chain
payment channels, \ie, a payment channel network~(PCN).
PCNs enable rapid payment
processing between any two channel endpoints 
without consulting the blockchain every time.
That is, incremental updates are negotiated locally instead of requiring a global
agreement. %
Local transactions are not only much faster, but also do not leak information to
noninvolved third parties. This local negotiation is however
only possible in a secure manner, because the parties deposit
a collateral during channel establishment.
In combination with a clever exchange of multi-signature transactions,
the blockchain serves as a mechanism to resolve conflicts.
In addition to payments between neighboring nodes,
the Lightning Network allows payments to traverse over multiple
channels in the network graph, \ie, multi-hop payments.
Therefore, channel balances have to be updated atomically,
which is enforced by the
utilization of Hashed Time-Locked Contract (HTLC) payment protocols.
In an attempt to avoid information leakage to intermediate nodes,
sender and receiver exchange messages by applying an onion routing
encryption scheme; more specifically, the Sphinx mix packet construction~\cite{danezis2009sphinx}.

In this work, we however show that the privacy guarantees of the Lightning
Network may be subverted by an adversary conducting timing attacks on the
message exchange during payment processing.
In particular, an on-path adversary may reduce the anonymity set of potential
sender and receiver nodes based on the payment amount and the HTLC's time-lock
delta value. Following this initial reduction of privacy, the adversary may
apply timing-based estimators to infer the likeliest payment path
endpoints, potentially deanonymizing the sender and receiver of a
payment. This attack is especially fatal, since countermeasures
directly conflict with the design goal of secure and rapid payments.
Moreover, as our analysis shows, the single most central node is already
capable of observing close to 50\% of all payments in the network,
while the four most central nodes observe an average of 72\% payments. These
findings are in accordance with recent results~\cite{tochner2019hijacking}
and emphasize the relevance of the on-path attacker model.

We expose that an adversary can probe the network and is able to derive a model of edge latencies,
which enables timing attacks. Furthermore, we show how the observation
of timing patterns, inherent to interactive multi-hop message exchanges, may be used by the
adversary to calculate time differences that correspond to her distance from the
respective payment endpoint. To this end, we introduce timing-based
estimators that first exclude invalid payment paths, before ranking candidate
nodes according to their likelihood, \ie, return a maximum likelihood estimation.
To confirm the feasibility of retrieving such measurements, we developed a
proof-of-concept implementation of the measurement functionality and deployed
it on a segregated part of the Lightning Network testnet. Furthermore, to
enable evaluation of the attack vector in larger scenarios, we
developed a discrete-event network simulator that allows to simulate the
payment routing protocol based on real-world snapshots of the public Lightning Network.
Utilizing the simulator, we study the timing attacks in scenarios
modelling adversarial capabilities of different magnitudes.
The results show that an adversary controlling more than 10 nodes
could easily deanonymize payment sources and destinations with a precision of more than 50\%.
The sensitivity highly depends on the malicious nodes' position in the network, though,
but can reach up to 50\% recall.
Moreover, we show that our time-based estimators are generally outperforming a
First-Spy estimator, which serves as a baseline.
This superior result becomes particularly apparent when considering full deanonymization,
\ie, the case in which the adversary was able to correctly identify both payment endpoints.

To summarize our contributions, we
(1) present a unified model for the Lightning Network that captures the payment channel graph as well as properties of the underlying peer-to-peer network,
(2) propose a method for probing the network to build an (adversarial) edge latency model,
(3) introduce timing attacks on privacy in payment channel networks that make
use of time difference measurements of interactive multi-hop message
exchanges, and
(4) analyze the feasibility and adversarial success of the introduced attack vector based on a proof-of-concept measurement node and comprehensive network simulations.

The remainder of this paper is structured as follows. Section~\ref{sec:ln}
gives a primer on the Lightning Network and its payment protocol.
Section~\ref{sec:model} introduces models and notations that serve as the
basis for our further analysis. In Section~\ref{sec:attack}, we introduce timing
attacks on privacy in payment channel networks, and evaluate their feasibility
and adversarial success in Section~\ref{sec:eval}. Section~\ref{sec:relwork} discusses related work,
before Section~\ref{sec:conclusion} concludes the paper.

\section{Lightning Network Primer}\label{sec:ln}
The Lightning Network is the most prevalent \emph{payment channel network
(PCN)} to date, \ie, it is a network of \emph{payment channels} that are
established between two endpoints by locking a certain amount of funds (the
channel \emph{capacity}) on-chain, whose individual allocation (the respective channel
\emph{balances}) can then be negotiated
rapidly between the two involved parties. Payments routed over multiple
intermediate channels allow to send money to only remotely connected receivers
while being secured through the application of \emph{Hashed Time-Locked
Contract (HTLC)} payment protocols.
The HTLC protocol ensures that a forwarding intermediary node is reimbursed in
the case of payment success, and in case of a failure may still retrieve its
locked funds after expiration of the \emph{time-lock delta} safety period.
In the following, we give a technical overview of the Lightning Network's channel
construction, routing, and payment processing mechanics.

\subsection{Connection and Channel Establishment}
A new peer joining the Lightning Network has first to
establish a network connection to a node connected to Lightning's TCP-based
peer-to-peer overlay network. Since every node in the network holds an
associated long-term \texttt{secp256k1}~\cite{certicom2010eccparameters}
public key by which it is identified, all inter-peer communications following
the initial key exchange handshake are authenticated and encrypted based on
the Noise~\cite{perrin2018noise} protocol framework. 

In order to initiate the establishment of a new payment channel to a 
neighboring node, the peer sends an \texttt{open\_channel} message that is
typically answered by an \texttt{accept\_channel} message, through both of which the
channels parameters, in particular the
channel capacity and initial balances, are negotiated. Using the
exchanged information, the initiating peer is then able to issue a funding
transaction which it broadcasts in the Bitcoin network. After the funding
transaction is confirmed on-chain, the channel is established and
may be used for payment processing. Furthermore, if the new peer wants to act as  payment
hub, \ie, forward payments for others, it can announce the node's and
channel's existence to the network by disseminating the respective
\texttt{node\_announcement} and \texttt{channel\_announcement}
messages in the peer-to-peer network. As these messages also contain the
necessary routing information, such as the channel capacity and associated routing
fees, they are broadcasted in Lightning's overlay network. These messages
also include the \texttt{cltv\_expiry\_delta} parameter, which allows a node
to declare the maximum time it is willing to have its funds locked up in case
an HTLC is not fulfilled in an orderly fashion.

\subsection{Payment Routing}
\begin{figure}
    \centering
    \tikzset{node distance=4cm, every node/.style={thick, font=\small\sffamily},
routenode/.style={shape=circle,draw,minimum size=1.2em, fill=sron0, draw=sron0, text=white},
channel/.style={ultra thick},
coords/.style={node distance=0.35cm},
message/.style={->, >=stealth'},
handshake/.style={->, >=stealth',gray},
invoice/.style={->, >=stealth', dashed, right},
messagelabel/.style={font=\footnotesize, fill=white, sloped},
handshakelabel/.style={font=\footnotesize, sloped, fill=white, text=gray},
}

\begin{tikzpicture}[]
    \node[routenode]      	(a)      	at (0,0)                {A};
    \node[routenode, right of=a]      	(b)     			 		{B};
    \node[routenode, right of=b]      	(c)     			 		{C};

    \coordinate[coords, below of=a] (h00) {};
    \coordinate[coords, below of=h00] (h01) {};
    \coordinate[coords, below of=h01] (h02) {};
    \coordinate[coords, below = 0.15cm of h02] (h02') {};
    \coordinate[coords, below of=h02'] (h03) {};
    \coordinate[coords, below of=h03] (h04) {};
    \coordinate[coords, below = 0.15cm of h04] (h04') {};
    \coordinate[coords, below of=h04'] (h05) {};
    \coordinate[coords, below of=h05] (h06) {};
    \coordinate[coords, below of=h06] (h07) {};
    \coordinate[coords, below = 0.15cm of h07] (h07') {};
    \coordinate[coords, below of=h07'] (h08) {};
    \coordinate[coords, below of=h08] (h09) {};
    \coordinate[coords, below = 0.15cm of h09] (h09') {};
    \coordinate[coords, below of=h09'] (h0A) {};
    \coordinate[coords, below = 0.15cm of h0A] (h0A') {};
    \coordinate[coords, below of=h0A'] (h0B) {};
    \coordinate[coords, below of=h0B] (h0C) {};
    \coordinate[coords, below of=h0C] (h0D) {};
    \coordinate[coords, below of=h0D] (h0E) {};
    \coordinate[coords, below of=h0E] (h0F) {};

    \coordinate[coords, below of=b] (h10) {};
    \coordinate[coords, below of=h10] (h11) {};
    \coordinate[coords, below of=h11] (h12) {};
    \coordinate[coords, below = 0.15cm of h12] (h12') {};
    \coordinate[coords, below of=h12'] (h13) {};
    \coordinate[coords, below of=h13] (h14) {};
    \coordinate[coords, below = 0.15cm of h14] (h14') {};
    \coordinate[coords, below of=h14'] (h15) {};
    \coordinate[coords, below of=h15] (h16) {};
    \coordinate[coords, below of=h16] (h17) {};
    \coordinate[coords, below = 0.15cm of h17] (h17') {};
    \coordinate[coords, below of=h17'] (h18) {};
    \coordinate[coords, below of=h18] (h19) {};
    \coordinate[coords, below = 0.15cm of h19] (h19') {};
    \coordinate[coords, below of=h19'] (h1A) {};
    \coordinate[coords, below = 0.15cm of h1A] (h1A') {};
    \coordinate[coords, below of=h1A'] (h1B) {};
    \coordinate[coords, below of=h1B] (h1C) {};
    \coordinate[coords, below of=h1C] (h1D) {};
    \coordinate[coords, below of=h1D] (h1E) {};
    \coordinate[coords, below of=h1E] (h1F) {};

    \coordinate[coords, below of=c] (h20) {};
    \coordinate[coords, below of=h20] (h21) {};
    \coordinate[coords, below of=h21] (h22) {};
    \coordinate[coords, below = 0.15cm of h22] (h22') {};
    \coordinate[coords, below of=h22'] (h23) {};
    \coordinate[coords, below of=h23] (h24) {};
    \coordinate[coords, below = 0.15cm of h24] (h24') {};
    \coordinate[coords, below of=h24'] (h25) {};
    \coordinate[coords, below of=h25] (h26) {};
    \coordinate[coords, below of=h26] (h27) {};
    \coordinate[coords, below = 0.15cm of h27] (h27') {};
    \coordinate[coords, below of=h27'] (h28) {};
    \coordinate[coords, below of=h28] (h29) {};
    \coordinate[coords, below = 0.15cm of h29] (h29') {};
    \coordinate[coords, below of=h29'] (h2A) {};
    \coordinate[coords, below = 0.15cm of h2A] (h2A') {};
    \coordinate[coords, below of=h2A'] (h2B) {};
    \coordinate[coords, below of=h2B] (h2C) {};
    \coordinate[coords, below of=h2C] (h2D) {};
    \coordinate[coords, below of=h2D] (h2E) {};
    \coordinate[coords, below of=h2E] (h2F) {};

    \draw[channel] (a)	-- 	(b)	{};
    \draw[channel] (b)	-- 	(c)	{};

    \draw[invoice] (c.north) -- +(0,0.5cm) -| node[messagelabel, pos=0.25,
    above] {invoice: $H(r)$}	(a)	{};

    \draw[] (a)	-- 	(h0D)	{};

    \draw[] (b)	-- 	(h1D)	{};

    \draw[] (c)	-- 	(h2D)	{};

    \draw[message] (h00) -- node[messagelabel] {\texttt{update\_add\_htlc}: $H(r),\circledcirc$} (h11);
    \draw[handshake] (h01) -- node[handshakelabel] {\texttt{commitment\_signed}} (h12);
    \draw[handshake] (h12') -- node[handshakelabel] {\texttt{revoke\_and\_ack}} (h03);
    \draw[handshake] (h13) -- node[handshakelabel] {\texttt{commitment\_signed}} (h04);
    \draw[handshake] (h04') -- node[handshakelabel] {\texttt{revoke\_and\_ack}} (h15);

    \draw[message] (h1B) -- node[messagelabel] {\texttt{update\_fulfill\_htlc}: $r$} (h0C);

    \draw[message] (h15) -- node[messagelabel] {\texttt{update\_add\_htlc}: $H(r),\circledcirc$} (h26);
    \draw[handshake] (h16) -- node[handshakelabel] {\texttt{commitment\_signed}} (h27);
    \draw[handshake] (h27') -- node[handshakelabel] {\texttt{revoke\_and\_ack}} (h18);
    \draw[handshake] (h28) -- node[handshakelabel] {\texttt{commitment\_signed}} (h19);
    \draw[handshake] (h19') -- node[handshakelabel] {\texttt{revoke\_and\_ack}} (h2A);
    \draw[message] (h2A') -- node[messagelabel] {\texttt{update\_fulfill\_htlc}: $r$} (h1B);
\end{tikzpicture}
    \caption{Message exchange during payment routing.}
    \label{fig:htlc}
\end{figure}
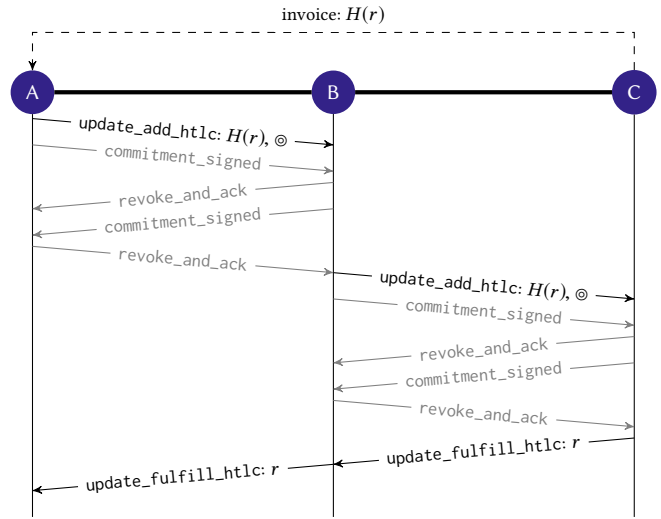

Let's assume that Alice already connected her node $A$ to the network and
established at least one channel over which she is able to send and receive payments. If
she now wants to send a payment to a destination node $C$, she has to first find a
suitable path in the network and then has to setup the corresponding HTLC to conduct the
payment. In order to illustrate this example, the sequence of exchanged
messages is shown in Figure~\ref{fig:htlc}.

Initially, it is required for $A$ and $C$ to have some out-of-bounds communication channel
over which $C$ can supply an \emph{invoice} to $A$ that includes, without
limitation, its identifying public
key, the amount to be paid, as well as the \emph{payment hash}, \ie, the hash $H(r)$ 
of a random secret $r$. Based on the publicly available routing information, $A$
then employs source routing in order to determine a path to $C$ that has
sufficient capacity to possibly be able to route her payment. While the behavior of
the source routing algorithm is not specified as part of the
BOLT specifications~\cite{lnrfc}, typically a modified version of Dijkstra's shortest path
algorithm~\cite{dijkstra1959note} that considers routing fees and past payment
success is utilized for route selection. Note that this algorithm might fail, if there
is no path of sufficient capacity available. However, let's assume without
loss of generality that this
is not the case and the algorithm yields a path over the intermediate node $B$. 

Given this path, Alice is able to initiate the HTLC construction, \ie,
a number of conditional payments that either may be redeemed by producing the
pre-image $r$ to the challenge $H(r)$ or would time out after a certain lock-time. 
In order to facilitate the payment, $A$ calculates two essential values for
each respective hop: 
\begin{enumerate}
    \item the amount this hop should forward,
        which is calculated by adding the accruing fees for each respective
        hop to the payment amount, which is hence \emph{increasing} towards
        the destination.
    \item the necessary remaining time-lock value for the outgoing hop, which is decreasing towards the
        destination.
\end{enumerate}

Alice then encodes this information in an onion routed packet
corresponding to the Sphinx packet scheme~\cite{danezis2009sphinx}. That is,
the packet is constructed through multiple layers of encryption, each wrapping
information identifying the next hop, the amount to forward, as well as the
remaining lock-time value. She then initiates the payment by sending a
\texttt{update\_add\_htlc} message carrying the payment hash $H(r)$ as well as the
onion packet (represented as~$\circledcirc$) to the next hop, \ie, $B$.
The latter, as each intermediate node along the payment path, is then able to decrypt the packet to
receive its payload and forward the HTLC offer to the next hop. However, $B$
only proceeds after the new conditional
payment based on the challenge $H(r)$ is is incorporated in the state of the
affected payment channel and this state change is \emph{irrevocably committed}
through a handshake of \texttt{commitment\_signed} and \texttt{revoke\_and\_ack} messages.

After the pending state updates have been negotiated, $C$ forwards the HTLC
by attaching the remainder of the onion packet to an \texttt{update\_add\_htlc} message sent to the next hop, which proceeds in the same way. 
In case any of the intermediary nodes does not agree with
the payment, \eg, when the channel does not hold a sufficient balance or fee and time-lock values determined $A$ do not meet their expectations, they may fail the HTLC by replying with an
\texttt{update\_fail\_htlc} message carrying failure
message that is onion-encrypted and propagated back along the path to the
origin node $A$. As this may happen at any point in time, Lightning does not
provide any guarantees on payment reliability and hence can be classified as a
best-effort network. 

Once the HTLC construction reaches the final destination, $C$ supplies the solution $r$ to the payment hash challenge $H(r)$ via a corresponding \texttt{update\_fulfill\_htlc} message, which is
propagated back on the inverse payment path, allowing intermediary nodes to redeem their
conditional payments. Thereby, they settle the pending HTLC and gain the
determined fee. Note that while the \texttt{commitment\_signed} and
\texttt{revoke\_and\_ack} messages are only exchanged between
immediate neighbors, the \texttt{update\_add\_htlc},
\texttt{update\_fulfill\_htlc}, and \texttt{update\_fail\_htlc} messages are forwarded
back and forth the payment path, which makes them observable by intermediate
nodes.

\section{Model}\label{sec:model}
In the following, we introduce the models and notations that serve as the
basis for further analysis.

\subsection{Network Model}
As discussed in the previous section, PCNs typically
exhibit multiple layers: while inter-peer communication is handled
by the peer-to-peer network layer, the payments themselves are sent and forwarded in the
network of payment channels. While peers may join the peer-to-peer
network without establishing payment channels, peers with an established
payment channel have to be connected in the peer-to-peer network. 
We in the following assume the peer-to-peer network to be congruent
with the channel layer and build a unified model based on the public network of payment channels.

A PCN can therefore be modeled as a single graph $\lngraphsymb =
(\lnverticessymb, \lnedgessymb, \phi)$, where
$\lnverticessymb = \{v_0, \ldots, v_n\}$ is the set of the network's nodes and
$\lnedgessymb = \{e_0, \ldots, e_m\}$ represent the set of edges, \ie, payment channels. Since every node may have multiple payment channels to any other
node, $\lngraphsymb$ is a loopless multigraph and 
$\phi: \lnedgessymb \rightarrow \{\{u,v\} \mid u,v \in \lnverticessymb \wedge u \neq v\}$
associates the set of edges with their endpoint nodes. 

In each direction, an edge $e$ is
associated with a \emph{balance} $\lnbal{e}{u}{v}$
that denotes the available balance from $u$ to $v$ on channel $e$, where
$u, v \in \phi(e)$. Edges also have an associated \emph{fee function},
defined by $\lnfee{e}{u}{v}{a}$ that
takes the payment amount $a$ as a parameter and yields the fees that accrue when
forwarding over this channel. Note that during routing directionality matters
and hence balance and fee functions are asymmetric. That is, generally
$\lnbal{e}{u}{v} \neq \lnbal{e}{v}{u}$ and $\lnfee{e}{u}{v}{a} \neq
\lnfee{e}{v}{u}{a}$. In
contrast, the edge \emph{capacity} is symmetric and defined as the sum of
balances, \ie, $\lncap{e} = \lnbal{e}{u}{v} + \lnbal{e}{v}{u}$. 
Furthermore, associated with the edges are the respective \emph{time-lock
delta} values $\lnlock{e}{u}{v}$ that indicate the maximum time in block
height the forwarding node is willing to have its funds locked in case an HTLC
fails. And lastly, the function $\lnlat{e}$ assigns a \emph{latency
distribution} to each
network edge that represents the network and processing delays,
which are induced when messages traverse the edge in the underlying peer-to-peer network.
Note that while balances and latencies are changing frequently and are only
known locally, fees, capacities, and time-lock requirements are considered
static and are publicly accessible by all nodes in order to enable the
source routing process. We denote this public graph information as $\lnpubgraphsymb = (\lngraphsymb,
\lncapsymb, \lnfeesymb, \lnlocksymb)$.

Based on this model of payment channel networks, we can now introduce the
following definitions.

\begin{definition}[Payment]
    A \emph{payment} is defined as the tuple 
    \begin{displaymath}
        (s,t,a,\lnmaxlocksymb),
    \end{displaymath}
    where $s$ and $t$ respectively denote the origin and destination nodes,
    $a$ is the amount sent by $s$, and $\lnmaxlocksymb$ signifies the
    maximal total time the payment amount may be locked.
\end{definition}

\begin{definition}[Path Validity]\label{dfn:validity}
    A \emph{path} from node $s$ to node $t$ in the network is a sequence of
    connecting edges, denoted as a tuple 
    \begin{displaymath}
        p = (e_0, e_1, \ldots, e_l), 
    \end{displaymath}
    where $\phi(e_0) = \{s, v_1\}, \phi(e_1) = \{v_1, v_2\}, \ldots, \phi(e_l) = \{v_{l}, t\}$.

    A path $p$ is called \emph{timelock-valid} w.r.t.\ a total time-lock delta
    $\lnmaxlocksymb$, if the remaining time the payment might be locked is
    smaller than the time the forwarding node would be willing to accept, \ie,
    \begin{displaymath}
        \forall e_i \in p,\,u_i,v_i\in \phi(e_i):\,\lnlock{e_i}{u_i}{v_i} \geq
        \lnmaxlocksymb - \sum\limits_{j=0}^{i-i} \lnlock{e_j}{u_j}{v_j} 
    \end{displaymath}

    Furthermore, a path $p$ is \emph{capacity-valid} w.r.t.\ an amount $a$, if
    all edges have capacities higher than the forwarding amount including accruing fee,
    \ie, 
    \begin{displaymath}
        \forall e_i \in p,\,u_i,v_i\in \phi(e_i):\,\lncap{e_i} \geq \mathbf{f}_i,
    \end{displaymath}
    where $\mathbf{f}_i$ is defined recursively as $\mathbf{f}_l = a$ and $\mathbf{f}_{i-1} =
    \mathbf{f_i} + \lnfee{e_i}{u_i}{v_i}{\mathbf{f}_i}$. Note that a path $p$ being
    capacity-valid for amount $a$ does not necessarily imply that a payment of size $a$
    can actually be routed over $p$.

    A payment can be routed only if the path does not only
    exhibit sufficient capacities, but also balances to forward the respective
    amount,
    \begin{displaymath}
        \forall e_i \in p,\,u_i,v_i\in \phi(e_i):\,\lnbal{e_i}{u_i}{v_i} \geq \mathbf{f}_i,
    \end{displaymath}
    in which case we call it \emph{balance-valid} or \emph{$a$-routable}.

    Independently of this special case, however, we call a payment path
    \emph{generally valid} or just \emph{valid}, if it is timelock-valid \emph{and}
    capacity valid. Note that therefore path validity describes if a path
    could potentially be used to route a
    payment with respect to the parameters, not if it actually may be used or is being
    used.
\end{definition}

\begin{definition}[Routing Algorithm]
    A \emph{routing algorithm} $\lnroutealgsymb$ is a function that takes a
    payment $x = (s,t,a,\lnmaxlocksymb)$ and the public graph information
    $\lnpubgraphsymb = (\lngraphsymb, \lncapsymb, \lnfeesymb, \lnlocksymb)$
    as arguments and outputs a valid payment path,
    \ie, 
    \begin{displaymath}
        \lnroutealg{\lnpubgraphsymb}{x} \rightarrow p,
    \end{displaymath}
    where $p$ is a capacity-valid and timelock-valid path from $s$ to $t$.
\end{definition}

\begin{definition}[Reachability]
    In the graph $\lngraphsymb$, a node $t$ is \emph{reachable} from a node
    $t$ if there is a path between them. Similarly, we call~$t$
    \emph{capacity-reachable},\emph{balance-reachable}, or \emph{timelock-reachable} from~$s$, if there
    exists respectively a capacity-valid, balance-valid, or timelock-valid
    path from~$s$ to~$t$, w.r.t.\ a given payment $(s,t,a,\lnmaxlocksymb)$.
    
    Note that these reachability notions induce subgraphs $\lnrcapsymb$,
    $\lnrbalsymb$, and $\lnrlocksymb$, where
    \begin{displaymath}
        \lngraphsymb \supseteq \lnrcapsymb \supset \lnrbalsymb \quad
        \text{and} \quad \lngraphsymb \supseteq \lnrlocksymb.
    \end{displaymath}
\end{definition}

\subsection{Adversary Model}

\subsubsection{Lightning's Security Goals}
Given payments are routed directly between source and
destination nodes, secured by the HTLC construction, and the path is obscured by
employing the Sphinx-based onion routing scheme, the Lightning Network
aims to deliver the following security goals:
\begin{description}
    \item[Balance security:] No third party should be able to steal funds, or
        otherwise alter channel balances without the implicit consent of the
        involved parties.
    \item[Off-path local unobservability:] Only nodes on the payment path
        should be informed about an occurring payment.\footnote{Note however that this
        assumes a local perspective on the network. Payment unobservability
        may not hold when we assume a more powerful attacker model, such as an
        adversary that has access to large parts of the underlying network
        infrastructure. Such adversaries are known to be potentially capable of
        advanced deanonymization attacks, and are notoriously hard to
        defeat.~\cite{johnson2013usersrouted}}
    \item[Off-path value privacy] Since all communications during payment
        processing are encrypted, only on-path nodes should get to know the amounts forwarded.
    \item[On-path sender/receiver-anonymity:] As every node only knows its
        predecessor and successor on the payment path, it should not be able
        to identify the sender or receiver of a payment. 
    \item[Receiver's sender-anonymity:] The receiver of a payment should
        not be able to identify who initiated a payment.
        \footnote{Notably, the Lightning Network currently does not guarantee the inverse, \ie,
        the possibility for a receiver to stay anonymous. However, this may
        feasible in the future, when the currently discussed \emph{Rendez-Vous
        Routing} proposal is implemented.~\cite{lnrfcrendez}}
\end{description}

\subsubsection{Adversarial Goals and Capabilities}
While the off-path unobservability of payments could potentially be subject of a deanonymization attack run by a
global passive adversary, in this work we analyze the feasibility of
subverting the on-path anonymity properties of nodes that send and receive payments.
In particular, we focus on attack vectors that allow a local
adversary incorporating side-channel information to potentially subvert \emph{on-path
sender/receiver anonymity} as well as \emph{receiver's sender-anonymity}. 

To this end, we assume an internal local adversary that controls a set
$M~=~\{m_0,~\ldots,~m_k\}$ of malicious nodes in the network that act as
payment-processing intermediaries, which in accordance with literature may
also be referred to as \emph{spies}. We furthermore
assume that the adversarial nodes $M$ behave according to protocol and are
able to send and receive protocol-compatible messages, \eg, in order to probe the network to build a latency
model of their surroundings. 

When payments are routed over an adversarial node $m_i \in M$, it keeps track of
each network message \texttt{msg} arriving over the edge $e_m$, as well as the corresponding timestamp, \ie, they store the
datasets 
\begin{displaymath}
    \lndatasymb_i = \{(m_i, e_m, t_\texttt{msg}, \texttt{msg})\}.
\end{displaymath}

Based on the merged dataset $\lndatasymb = \bigcup_i \lndatasymb_i$, the
public graph data $\lnpubgraphsymb$, 
and the estimated link latencies $\widehat\lnlatsymb$, the adversary then
aims to associate any observed payments $x = (s,t,a,\lnmaxlocksymb)$ with the
respective source node $s$ and destination node $t$. For this classification, the adversary may
apply different \emph{source and destination estimators} $\lnestimatorsymb_s$ and
$\lnestimatorsymb_t$ that given the input data yield a respective estimation,
\ie,
\begin{displaymath}
    \lnestimator{\lndatasymb}{\lngraphsymb_{pub}}{\widehat\lnlatsymb}{s}{x} =
    \widehat{v}_s \quad\text
    {and} \quad
    \lnestimator{\lndatasymb}{\lngraphsymb_{pub}}{\widehat\lnlatsymb}{t}{x} =
    \widehat{v}_t,
\end{displaymath}
where $\widehat{v}_s, \widehat{v}_t \in \lnverticessymb$. For the sake of brevity, we in the
following refrain from always giving an exhaustive list of arguments and opt
to abbreviate notation as $\lnestimatorsymb_s(x|\lndatasymb)$ and $\lnestimatorsymb_t(x|\lndatasymb)$.

\subsection{Anonymity Metrics}
In order to quantify adversarial success and analyze the privacy properties of
the network, we utilize the following privacy metrics.

Well known performance measures for the adversarial success of estimator-based
deanonymization attacks are the combination of \emph{precision} and
\emph{recall}~\cite{fanti2018dandelion++}.

Assuming $X$ being the set of all payments and $C \subseteq X$ the set of all
payments observed and classified by the adversary. Let furthermore $X_u
\subseteq X$ denote the set of all payments that originate from (or end at, in case of
destination estimation) node $u$ and analogously $C_u$ denote the set of
payments classified to originate from (end at) node $u$, \ie, 
\begin{displaymath}
C_u = \{x \mid \lnestimatorsymb(x|\lndatasymb) = u\}. 
\end{displaymath}

Then the \emph{precision} $D$ of the estimator $\lnverticessymb$ is defined as
the share of classified payments that were indeed correctly classified, \ie,
\begin{displaymath}
    D = \frac{|C_u \cap X_u|}{|C|}.
\end{displaymath}

The estimator's \emph{recall} $R$ however is the share of all payments in the
network that were correctly classified,
\begin{displaymath}
    R = \frac{|C_u \cap X_u|}{|X|}.
\end{displaymath}

A unified measure for the accuracy of an estimator is given by the harmonic mean of
precision and recall, also known as the $F_1$-measure: 
\begin{displaymath}
    F_1 = 2 \cdot \frac{D \cdot R}{D + R}.
\end{displaymath}

\section{Timing Attacks on Privacy}\label{sec:attack}
In the following, we describe the steps necessary to conduct timing attacks on
privacy in payment channel networks.

\subsection{Improving Topological Advantage}\label{sec:attack:topoadv}
The attacker wants to maximize the number of payment paths it is included
in by the victim's routing algorithm. While the client-side routing
behavior is not standardized as part of Lightning's BOLT specifications, most implementations of
the Lightning protocol rely on modified versions of Dijkstra's shortest path
algorithm~\cite{dijkstra1959note} that consider the channel fees as well
as other parameters. Note that, as recent
literature has observed~\cite{mizrahi2020congestion}, more than 90\% of
today's Lightning Network nodes run the \texttt{LND} implementation of the
Lightning protocol. In the following, we therefore assume \texttt{LND} to
be the default implementation and use it as the base of our further
analysis. \texttt{LND}'s path finding algorithm 
selects candidate edges based on a weight function $w_e$ that considers routing
fees for the routed amount $a$, as well as a risk factor $r_f$ that aims
to capture the worst-case lock time: 
\begin{equation*}
    w_e = \lnfee{a}{e}{u}{v} + a \cdot \lnlock{e}{u}{v} \cdot r_f,
\end{equation*}
where $r_f = 1.5 \cdot 10^{-8}$ is the default configuration.

Therefore, by setting time lock parameters and channel fees to the minimal
allowed values, the adversary can minimize the routing weight function of the
victim's client software, and thereby maximize the probability that at
least one of the malicious nodes is included in the payment path. 
Tochner et al.~\cite{tochner2019hijacking}
studied this kind of \emph{route hijacking} in the
context of denial-of-service attacks. They showed that
at the time of writing ten nodes are part of 80\% of all payment paths,
and 30 nodes of over 95\% of payment paths.
Furthermore, while the problem of optimal edge additions for maximum betweenness
centrality has previously been show to be
NP-hard~\cite{bergamini2018improving, avarikioti2019ride,
ersoy2019profit}, the authors provide a greedy algorithm with which
an adversary improve its topological advantage. To this end, they were
able to show that the creation of only fifteen edges would suffice to
hijack more than 80\% of \texttt{LND} payment paths.
Their observations are generally in accordance with our findings regarding adversarial
path inclusion (cf.\ Section~\ref{sec:eval:pathinclusion}) and 
highlight the relevance of the on-path attacker model.

\subsection{Building the Latency Model}\label{att:timingmodel}
As a data basis for the classification of observed
payments, the adversary initially has to probe the network to retrieve characteristic timing measurements. 
These measurements allow her to build a model of latencies $\widehat\lnlatsymb$ that are encountered when payments are routed over a specific
link, which then in turn are used as a priori knowledge for the estimators.

\subsubsection{Retrieving Path Latency Measurements}
In order to probe for the characteristic latency measurements, the adversary
can exploit the fact that due to Lightning's use of the Sphinx packet format,
invalid or failing payments can only be discovered by the node that they are
actually failing at. That is, as all nodes only see the parts of onion-routed data
they are able to encrypt and do not know the full path's properties, they have
to optimistically forward all payment requests based on the assumption that it
will succeed. Therefore, the adversary is able to craft payments that look
valid to all intermediaries, but are bound to fail at a specific hop along the
path, \eg, because of insufficient fees or an invalid maximum time-lock value. 
Utilizing this probing method, the adversary can record the time difference
between sending the initial \texttt{update\_add\_htlc} message and retrieving the
final \texttt{update\_fail\_htlc} to retrieve a measurement that encompasses
all delays that were encountered along the measured payment path.

\subsubsection{Estimating Edge Latencies}
The adversarial node utilizes the described probing method to 
retrieve a reliable model for paths covering every link in the network. To
this end, she iteratively increases the probing path lengths and calculates link latencies
by subtracting the estimated latencies of partial paths. That is, the adversarial node
$m_i$ starts by repeatedly probing paths lengths $l=1$ that cover its immediate
neighbors $v_j$, \ie, $p_1=(e_1), m_i, v_j \in \phi(e_1)$, and calculates the mean
$\mu_{e_1}$ and standard deviation $\sigma_{e_1}$ values for these links, \ie, 
\begin{equation*}
    \widehat{\mu}_{e_1} = \frac{\sum_{i=0}^{n} \texttt{probe}_i(p_1)}{T\cdot n},
\end{equation*}
\begin{equation*}
    \widehat{\sigma}_{e_1} = \sqrt{\frac{\sum_{i=0}^{n} (\texttt{probe}_i(p_1) - \widehat{\mu}_{e_1})^2}{T\cdot n}},
\end{equation*}
where $T$ is a normalizing factor accounting for the number of link traversals
incurred during the message exchange over the measured hop. In this case, we
assume $T = 4$, \ie, three traversals for $\texttt{commitment\_signed}$ and
for the $\texttt{revoke\_and\_ack}$ handshake and one for $\texttt{update\_fulfill\_htlc}$ (see Figure~\ref{fig:htlc}).

The adversary can then increase the path lengths and iteratively build the
latency model for these longer paths $p_l = (e_1, \ldots, e_l)$:
\begin{equation*}
    \widehat{\mu}_{e_l} = \frac{\sum_{i=0}^{n} \texttt{probe}_i(p_l)}{T\cdot
    n} - \widehat{\mu}_{e_{l-1}} - \ldots - \widehat{\mu}_{e_1},
\end{equation*}
\begin{equation*}
    \widehat{\sigma}_{e_l} = \sqrt{\frac{\sum_{i=0}^{n} (\texttt{probe}_i(p_l)
    - \widehat{\mu}_{e_1})^2}{T\cdot n} + ... + \widehat{\sigma}_{e_{l-1}}^2 +
    \widehat{\sigma}_{e_{l}}^2}.
\end{equation*}

Given these parameters, the adversary can build the normally distributed
edge latency model as 
\begin{equation*}
    \widehat{\lnlatsymb}(e_i) = \mathcal{N}(\widehat{\mu}_{e_i},
    \widehat{\sigma}_{e_i}^2).
\end{equation*}

Note that this model does not just include the network delay, but also
incorporates any processing delays arising on the intermediate nodes. As this
unified latency model captures various side effects, we can refrain from
considering them separately in the attack estimators.

Moreover, modeling timing behavior in such an approximative way is bound to induce a certain
margin of error. This uncertainty is expressed by the variances growing with
increasing lengths of the measured paths.
In the following we therefore propose a method to aggregate the timing models from
multiple malicious vantage points to increase overall accuracy.

\subsubsection{Model Aggregation}
As the adversary may control multiple nodes in the network to increase the
probability of inclusion in payment paths, each malicious node may create a
timing model from their point of view. As the margin of error increases with 
each additional hop in the measured paths, the aggregated model should not simply average over all
measurements. Instead, it merges the individual model by applying an
arithmetic mean weighted with the reciprocal distance from the measured node, \ie, 
\begin{equation*}
    \forall m_i \in M, v_i \in V: w_i = \frac{1}{d(v_i, m_i)},
\end{equation*}
\begin{equation*}
    \widehat{\mu}_{e, tot} = \frac{\sum_{i=0}^{n\cdot |M|} w_i \widehat{\mu}_{i, e}}{\sum_{i=0}^{n\cdot |M|} w_i},
\end{equation*}
\begin{equation*}
    \widehat{\sigma}_{e, tot} = \sqrt{\frac{\sum_{i=0}^{n\cdot |M|} w_i
    (\widehat{\mu}_{i, e}-\widehat{\mu}_{e, tot})^2}{\sum_{i=0}^{n\cdot |M|}
    w_i}}.
\end{equation*}

The adversary therefore retrieves the aggregated latency model
\begin{equation*}
    \widehat{\lnlatsymb}_{tot}(e_i) = \mathcal{N}(\widehat{\mu}_{e,tot}, \widehat{\sigma}_{e,tot}^2).
\end{equation*}

\subsection{Estimator-based Deanonymization Attack}
\begin{figure}
    \centering
    \tikzset{node distance=2.5em, every node/.style={thick, font=\small\sffamily},
routenode/.style={shape=circle,draw,minimum size=0.6em, fill=sron0, draw=sron0},
malicious/.style={preaction={fill, sron2}, draw=sron2, fill=sron2,
pattern=north west lines, pattern color=white},
channel/.style={ultra thick},
message/.style={->, >=stealth'},
}

\begin{tikzpicture}[]
    \node[routenode]      	(a)      	at (0,0)                {};
    \node[routenode, malicious,  below right of=a]      	(b)      	{};
    \node[routenode, malicious, above right=1em and 3em of a]      	(c)      	{};
    \node[routenode, below right=2em and 1em of c]      	(d)      	{};
    \node[routenode, malicious, right=3em of c]      	(e)      	{};
    \node[routenode, above right=2em and 2em of c]      	(f)      	{};
    \node[routenode, above right=3em and 0.5 em of a]      	(g)      	{};
    \node[routenode, right=3em of d, double, label=below:{t}]      	(h)      	{};

    \node[routenode, label=above:{s}, left= of a, double]      	(i)      	{};
    \node[routenode, below left= of a]      	(k)      	{};

    \node[routenode, below left=1em and 2.5em of i]      	(l)      	{};

    \node[routenode, above right=1em and 1 em of f]      	(m)      	{};
    \node[routenode, above left=1em and 1 em of f]      	(n)      	{};
    \node[routenode, above left=1em and 1 em of g]      	(o)      	{};
    \node[routenode, below right=0.5em and 1 em of b]      	(p)      	{};

    \draw[channel] (a) -- (b)	{};
    \draw[channel] (a) -- (c)	{};
    \draw[channel] (a) -- (g)	{};
    \draw[channel] (a) -- (i)	{};
    \draw[channel] (a) -- (k)	{};

    \draw[channel] (b) -- (c)	{};
    \draw[channel] (b) -- (d)	{};
    \draw[channel] (b) -- (k)	{};
    \draw[channel, malicious] (c) -- (e)	{};
    \draw[channel] (c) -- (d)	{};
    \draw[channel] (c) -- (f)	{};
    \draw[channel] (c) -- (g)	{};
    \draw[channel] (d) -- (h)	{};
    \draw[channel] (e) -- (f)	{};
    \draw[channel] (e) -- (h)	{};
    \draw[channel] (f) -- (m)	{};
    \draw[channel] (f) -- (n)	{};
    \draw[channel] (f) -- (g)	{};
    \draw[channel] (i) -- (l)	{};
    \draw[channel] (g) -- (o)	{};
    \draw[channel] (p) -- (b)	{};
    \draw[channel] (p) -- (d)	{};

    \draw[] (i) edge[message, bend left] (a);
    \draw[] (a) edge[message, bend left] (c);
    \draw[] (c) edge[message, bend left] (e);
    \draw[] (e) edge[message, bend left] (h);
\end{tikzpicture}
    \caption{Payment routed over malicious observation points.}
    \label{fig:network-attack}
\end{figure}
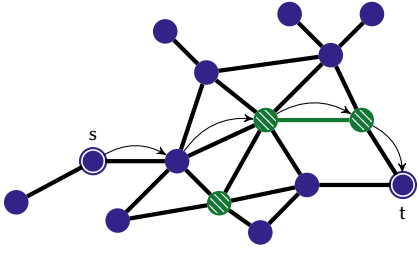

In order to be able to deanonymize the sender and receiver of a payment, 
it has to be routed over at least one observation point controlled by the
adversary (see Figure~\ref{fig:network-attack}).
In contrast to previous approaches that apply a \emph{First-Spy} estimator
that simply estimates the node adjacent to the point of observation to be the
payment's respective endpoint, our approach builds a maximum likelihood
estimator (MLE) over all paths the observed payment could possibly have taken.
To this end, the malicious nodes record the
time differences of interactive message exchanges and, after reducing the candidate set
by considering only valid paths (see Definition~\ref{dfn:validity}), estimate the source or destination of the payment according to the likelihood that the time differences stem from a
message exchange over this particular path.

\subsubsection{Recording Time Differences}
To utilize the timing of messages in order to estimate the source or
destination of a payment, the adversarial nodes $M$ have to observe end-to-end transmitted messages belonging to the
same payment at two different points in time, \ie, $t_0$ and  $t_1$. This allows
to calculate the time difference $\delta_t = t_1 - t_0$ it took the
observed payment to travel from the first point of observation $m_0 \in M$ to the
source or destination, and back to the second point of observation $m_1 \in
M$.

In particular, the malicious intermediate nodes record the point in time
$t_0$ when they forward a payment via \texttt{update\_add\_htlc} and $t_1$
upon receipt of the corresponding \texttt{update\_fulfill\_htlc}, which yields
a time difference $\delta_t$ corresponding to the distance to the payment's
destination (cf.~Figure~\ref{fig:htlc}). In this case, the adversary does not
have to interfere with the payment processing protocol in order to collect the necessary information 
to conduct destination estimation. Hence, the adversary acts in a purely 
honest-but-curious model, and therefore cannot be detected by outside parties.

However, since the message exchange from the source node to the intermediate
node is non-interactive, an advanced attack strategy is required in order to retrieve
suitable timing measurements in this direction. To this end, the adversary
intentionally fails the first observed payment attempt by sending a \texttt{update\_fail\_message}.
She also records the current time as $t_0$.
After receiving the failure message, the
payment's sender is forced to retry the failed attempt, which is typically done immediately
to avoid further delays. When the second payment attempt is
observed at time $t_1$, the adversary can calculate the $\delta_t = t_1 - t_0$
value which corresponds to its distance from the source node.

In general, the chosen paths and points of observation may be
different, in which case the adversary has only a certain chance of observing the second payment
attempt. While this would introduce additional uncertainty to this part of the
adversarial strategy, as we discuss later in Section~\ref{sec:eval:poc}, 
the adversary may force a sender to send the second payment attempt over the same path as
before, which removes this uncertainty.
This is possible in practice due to an implementation detail of \texttt{LND}.

We therefore in the following assume the two observations to occur at the same malicious node, \ie, $m_0 = m_1$.
Moreover, in the case that multiple malicious nodes are part of the payment
path and observe the payment, the
source or destination estimation is based on the measurements recorded
by the malicious node closest to the respective endpoint.

\subsubsection{Source and Destination Estimation}
\begin{algorithm}[t]
\caption{Source / Destination Estimator}
\label{alg:estimator}
\SetKwProg{estimate}{Function \emph{estimate}}{}{end}

\estimate{$\delta_t$, $a_{obs}$, $\Delta_{obs}$, $e_{obs}$, $\lnpubgraphsymb$}{
    remove all capacity-invalid paths from $\lnpubgraphsymb$

    remove all timelock-invalid paths from $\lnpubgraphsymb$

    \ForAll{$v \in V$} {
        initialize
    }

    $v_{fst}$ $\leftarrow$ \texttt{get\_neighbor($e_{obs}$)}
    \tcp*[f]{First hop is known}

    \texttt{queue\_candidate($v_{fst}$)}

    $\widehat{\lnlatsymb}_{fst}$ $\leftarrow$ $T_{obs} \cdot \widehat{\lnlatsymb}(e_{obs})$

    path\_lats[$v_{fst}$] $\leftarrow$ $\{\widehat{\lnlatsymb}(e_{obs})\}$

    likelihood[$v_{fst}$] $\leftarrow$ $\widehat{\lnlatsymb}_{fst} (\delta_t)$ 

    \While{$v_{cur}$ $\leftarrow$ \texttt{next\_unvisited()}} {
        $D_{cur}$ $\leftarrow$ path\_lats[$v_{cur}$]

        $\widehat{\lnlatsymb}_{cur} \leftarrow \sum\limits_{d_i \in D_{cur}} T_i \cdot d_i$
        \tcp*[f]{Aggregate dists.}

        $p_{cur}$ $\leftarrow$ $\widehat{\lnlatsymb}_{cur} (\delta_t)$

        \ForAll{$v_n$ in \texttt{neighbors($v_{cur}$)}} {
            $e_n$ $\leftarrow$ \texttt{cheapest\_edge($v_{cur}$, $v_n$)}

            $D_{n}$ $\leftarrow$ $D_{cur} \cup \{\widehat{\lnlatsymb}(e_n)\}$

            $\widehat{\lnlatsymb}_{n} \leftarrow \sum\limits_{d_i \in D_{n}} T_i \cdot d_i$

            $p_{n}$ $\leftarrow$ $\widehat{\lnlatsymb}_{n} (\delta_t)$

            $p_{old}$ $\leftarrow$ $likelihood[v_n]$
            
            \If{$p_{n} \leq p_{cur}$ {\bf or} $p_n \leq p_{old}$}{
                {\bf skip}
                \tcp*[f]{Only increasing likelihood}
            }

            likelihood[$v_{n}$] $\leftarrow$ $p_n$ \tcp*[f]{Update candidate}

            path\_lats[$v_{n}$] $\leftarrow$ $D_{n}$

            \texttt{queue\_candidate($v$)}
        }
    }
    \ForAll{visited $v$}{
    return candidate with max. likelihood
    }
}

\end{algorithm}

The estimation of source and destination of a payment relies on selecting the
likeliest paths the
payment could have taken before it arrived at the observation points. Therefore, in order to reduce the initial uncertainty, the adversary excludes paths that are capacity-unreachable or time-lock
unreachable given the observed amount $a_{obs}$ and $\Delta_{obs}$, \ie, she only
considers nodes in
\begin{displaymath}
    \lnrcapsymb \cap \lnrlocksymb \subseteq \lnpubgraphsymb .
\end{displaymath}

The adversary then builds candidate aggregated latency distributions
\begin{displaymath}
    \widehat{\lnlatsymb}_{p} = T_{obs}\cdot\widehat{\lnlatsymb}(e_{obs}) + ... + T_l \cdot \widehat{\lnlatsymb}(e_l)
\end{displaymath}
for each candidate path $p = (e_{obs}, ..., e_l)$, where $e_{obs}$ denotes the
edge the measurement was conducted through. Furthermore, the weights $T_i$ denote 
the number of messages that would have been exchanged over the edge $e_i$.
Note that the possibility of such an aggregation relies on the fact that the sum of normally distributed
variables may be calculated as
\begin{displaymath}
  \mathcal{N}(\mu_1, \sigma_1^2) + \mathcal{N}(\mu_2, \sigma_2^2) = \mathcal{N}(\mu_1 + \mu_2, \sigma_1^2 + \sigma_2^2).
\end{displaymath}

Then, the adversary ranks all candidate paths according to the likelihood that
the observed time difference $\delta_t$ was drawn from the respective
aggregated distribution, $\max_p(\widehat{\lnlatsymb}_p(\delta_t))$, and
estimates the final hop of the path to be the
payment's source or destination.

Therefore, the adversary generally would have to rank all possible paths in the network.
However, Algorithm~\ref{alg:estimator} implements the estimators
$\lnestimatorsymb_s$ / $\lnestimatorsymb_t$ as an
iterative algorithm that traverses the graph starting from the point of
observation. During execution, it adds new candidate paths as long as they would result in an
increased likelihood of observing $\delta_t$, and stops when all candidate
paths have been visited. 

\section{Evaluation}\label{sec:eval}
In the following, we evaluate the feasibility, accuracy, and reliability of
the presented attacks on privacy in payment channel networks.

\subsection{Ethical Considerations}\label{sec:ethics}
Research on the security and privacy of live communication systems is always in danger of
infringing on the rights of the participating individuals. In
accordance with the Menlo Report~\cite{bailey2012menloreport}, we
aim to minimize our interference with the live network as well as the data collected from unknowing parties.

That is, in order to evaluate the presented attacks on privacy in payment channel
networks, we pursue a two-pronged strategy: First we show the feasibility of the
attacks through a proof-of-concept implementation that was installed on an
entirely segregated part of the Lightning Network \texttt{testnet}, which ensures that no involuntary parties were
affected by our experiments. 

Second, to be able to evaluate larger attack scenarios and analyze the effect these attacks have
on the network's privacy overall, we rely on model-based network simulation that is not
connected in any way to unknowing individuals and hence does not
raise any ethical concerns. In particular, while the simulations utilize
latency measurements that were retrieved
through \emph{external} means, \ie, ICMP \texttt{ping} on nodes from the
public internet, we explicitly refrain from conducting \emph{internal}
latency measurements as discussed in Section~\ref{att:timingmodel}, since such
measurements could interfere with the functionality of the network.

\subsection{Proof-of-Concept Implementation}\label{sec:eval:poc}
The described attacks on payment privacy in payment channel networks rely on the
ability of malicious intermediary nodes to retrieve latency measurements from
the source and to the destination of an observed payment. As a proof of
concept that obtaining these measurements is indeed practical,
we implemented a respective plugin for
\texttt{c-lightning}~\cite{clightning}.\footnote{\label{fn:companion}Proof-of-concept
and simulator source codes, as
well as utilized data sets, are publicly available
in our companion repository at
\url{https://gitlab.tu-berlin.de/rohrer/cdt-data}}

\subsubsection{Retrieving Latencies to Destination}
When the plugin is started on the intermediary node, it registers to be
notified of forwarded payments. In particular, it utilizes the
\texttt{forward\_event} notification to record the times $t_0$ HTLC payment hashes
$H(r)$
are first observed, as well as the times $t_1$ they are marked as resolved. The
plugin furthermore records the node identifier $v_M$ of the measurement node and the
identifier $e_{next}$ of the channel the payment was forwarded over. That is, it records
the tuple $(H(r), u_M, e_{next}, t_0, t_1)$ that is then ready to be used as input
a payment destination estimator.

\subsubsection{Retrieving Latencies from Source}
Because the communication with the payment source is non-interactive, the
adversary has to rely on observing retried payment attempts, as discussed
above.  To this end, the proof-of-concept implementation makes use of the
\texttt{htlc\_accepted} hook provided \texttt{c-lightning}'s plugin API in order
to intercept incoming \texttt{update\_add\_htlc} messages. When a payment with
a previously unobserved payment hash $H(r)$ is observed, the plugin records
a corresponding timestamp $t_0$ and rejects the payment attempt. As the payment is
is then retried, the timestamp $t_1$ of the second observation by an
adversarial intermediary node is recorded. This hence allows the adversary to
estimate the latency from the source and record the tuple $(H(r), u_M,
e_{prev}, t_0, t_1)$.

While it is not guaranteed to observe the second payment, 
the proof-of-concept implementation is able to force the payment source to reuse the
same payment path by exploiting a weakness in the interplay of Lightning's
network protocol and \texttt{LND}-specific application behavior. That is, as channel
updates may occur in the middle of a payment attempt, \texttt{LND} elects not
to penalize intermediary nodes during route selection, if they report a channel
policy failure, \ie, fail the payment with
\texttt{amount\_below\_minimum}, \texttt{fee\_insufficient},
\texttt{incorrect\_cltv\_expiry}, or
\texttt{channel\_disabled} failure codes~\cite{lnbolt04, lndpolicyfailure}.
Note that \texttt{LND} once a minute grants such nodes a \enquote{second
chance}, independently of whether the returned channel policies entail an actually
meaningful update~\cite{lndsecondchancecommit}. This allows our
plugin to fail the first observed payment attempt with a corresponding
\texttt{update\_fail\_htlc} failure message, which prompts the \texttt{LND}
endpoint to immediately retry the payment over the same malicious intermediary
node, effectively enabling reliable latency measurements.\footnote{Note that
as of this writing, a small change in the \texttt{c-lightning}
source code is necessary to enable a plugin to return failure codes entailing
a channel policy update. A corresponding patch can be found in our companion
repository.}  

\subsubsection{Experimental Testnet Setup}
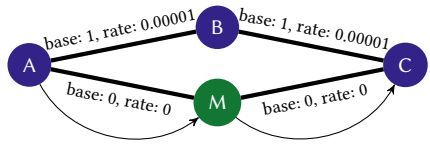
\begin{figure}
    \centering
    \tikzset{node distance=4cm, every node/.style={thick, font=\small\sffamily},
routenode/.style={shape=circle,draw,minimum size=1.2em, fill=sron0, draw=sron0, text=white},
evil/.style={sron2, text=white},
channel/.style={ultra thick},
message/.style={->, >=stealth'},
label/.style={font=\footnotesize, sloped},
}

\begin{tikzpicture}[]
    \node[routenode]      	(a)      	at (0,0)                {A};
    \node[routenode]      	(b)     	at (2.5,0.5)		 		{B};
    \node[routenode]      	(c)     	at (5,0)		 		{C};
    \node[routenode,evil]      	(m)     	at (2.5,-0.5)		 		{M};

    \draw[channel] (a)	-- node[label, above] {base: 1, rate: 0.00001}	(b)	{};
    \draw[channel] (a)	-- node[label, below] {base: 0, rate: 0} (m)	{};
    \draw[channel] (b)	-- node[label, above] {base: 1, rate: 0.00001}	(c)	{};
    \draw[channel] (c)	-- node[label, below] {base: 0, rate: 0}	(m)	{};

    \draw[] (a) edge[message, bend right=50] (m);
    \draw[] (m) edge[message, bend right=50] (c);

\end{tikzpicture}
    \caption{Experimental Testnet Setup}
    \label{fig:testnet}
\end{figure}

In order to confirm the feasibility of retrieving the required time
differences, we deployed an
experimental setup on a segregated part of Lightning's \texttt{testnet} network.
As shown in Figure~\ref{fig:testnet}, we deployed three nodes $A, B, C$ running
\texttt{LND} and one malicious node $M$ running \texttt{c-lightning} with our proof-of-concept
plugin. Between these nodes, channels were created so that the source node $A$ would have
two possible paths to send payments to destination node $C$: one over the benign
node $B$, and one over $M$. While the channels between the benign nodes were
configured with default fee settings ($\texttt{base\_fee} = 1$ and
$\texttt{fee\_rate} = 0.00001$), the malicious $M$ set its channel fees to
$0$ to increase its probability of payment path inclusion. 

We then sent payments in one minute intervals from node $A$ to node
$C$. For all payments, node $A$ chose the path $(A,M,C)$, which proves $M$'s
strategy to be successful. Moreover, as discussed above, $M$ would in each
case reject the first payment attempt and only proceed on the second try, allowing it to
retrieve latency measurements for both source node $A$
as well as destination node $C$.
It therefore confirms that we can retrieve the time differences
that pose the basis for our timing attacks.
As the next step, we can use the measurements to feed our estimators,
which would infer source and destination.

\subsection{Network Simulations}\label{sec:eval:netsim}
\subsubsection{Simulator \& Simulation Model}
In order to enable a larger-scale evaluation of the feasibility and impact of
timing attacks on privacy, we developed a network simulator that
allows to simulate payment routing in the Lightning Network based on
real-world data.\textsuperscript{\ref{fn:companion}}

The simulator consists of around 3,000 lines of Rust code that implement the network model
introduced in Section~\ref{sec:model}, as well as the logic to
run time-discrete simulations of multi-hop payments.
To this end, it recreates the multigraph of network nodes and edges as well as 
the necessary associated data (such as capacities, balances, time-lock deltas,
etc.) from a network snapshot.
Each node can queue events in simulation time,
\ie, a monotonically increasing clock with a resolution of 1\,ns.
This allows to simulate message exchange
according to times sampled from the underlying latency model, without
introducing unnecessary side-effects, even when the events happen
concurrently.
The messaging logic mimics the Lightning payment
protocol, making it possible to simulate and measure time differences in the
message exchange, \eg, as depicted in Figure~\ref{fig:htlc}. In order to find
payment paths, the simulator adopts the weight-based variant of Dijkstra's
algorithm from the \texttt{LND} implementation (see
Section~\ref{sec:attack:topoadv}).

The latency model is based on a measurement study
conducted in the Lightning network in March 2020. In this study, we retrieved
ICMP \texttt{ping} measurements to each reachable IPv4 address in the Lightning Network from various
geographically distributed vantage points. For more details, we refer the
reader to Appendix~\ref{app:lnp2p}. While initializing the graph model, the
simulator assigns a latency distribution to each edge based on the respective
geographic regions of the connected nodes. In case a node only advertises a
\texttt{.onion} address, \ie, is run behind a Tor hidden service, a random
geographic location is assigned.

For the following analysis, the simulator was parametrized with a snapshot of
the Lightning Network that was retrieved on May 1, 2020. Initial balance
distributions between channel endpoints were assumed to be a 50/50 split of
channel capacities. If not stated otherwise, in each simulated scenario 1,000 payments of varying amounts were sent between random
network nodes, and each scenario was repeated 30 times with different seed values
for the simulator's random number generator to ensure stastistical
significance.

In the following, we are considering three main adversarial scenarios: \texttt{mcentral},
\texttt{mrandom}, and \texttt{lnbig}. While in the \texttt{mcentral} case the
$m$ highest ranked nodes with respect to their \emph{betweenness centrality} are under control of the adversary, \texttt{mrandom}
acts as a baseline in which she only controls $m$ nodes chosen by uniform
random sampling. A special case is the \texttt{lnbig} scenario, in which we
study the potential capabilities of the 26 high-capacity nodes controlled by
the single entity \enquote{LNBIG.com}.

\subsubsection{Share of Compromised Paths}\label{sec:eval:pathinclusion}
\begin{figure}
    \centering
    \input{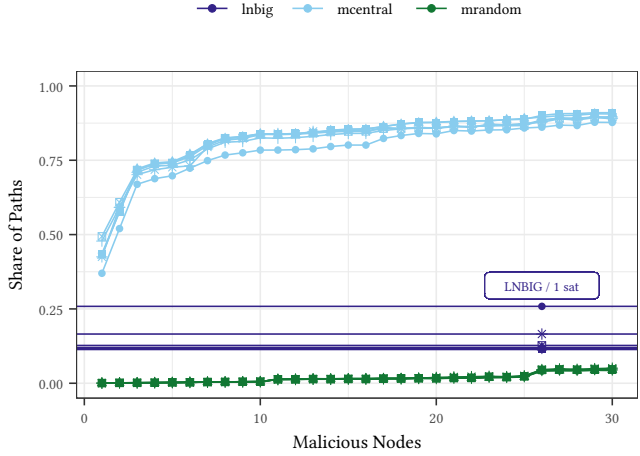}
    \caption{Share of compromised payment paths.}
    \label{fig:plot_path_inclusion}
\end{figure}
In order to evaluate the relevance of the on-path adversary model, we analyze
how likely it is that payments may be observed by adversaries of different magnitudes.
In each network scenario, we simulated payments of different amounts (1, 10,
100, 1,000, 10,000, and 100,000 satoshis) between randomly
chosen nodes and counted the times a malicious node was part
of the path returned by the routing algorithm. 

Figure~\ref{fig:plot_path_inclusion} shows the share of compromised paths 
for network scenarios in which the adversary controls the $m \in \{1, .., 30\}$ most central or random
nodes, as well as for the \texttt{lnbig} scenario. As this corresponds to the
definition of betweenness centrality, it comes to no surprise that the most
central nodes observe a high and increasing number of paths. However, it is
noteworthy that the single most central node is included in 37\% to around
49\% of payment paths, depending on the chosen amount. Additionally, the share
of compromised paths follows an initial steep increase, allowing an adversary in control of the
four most central nodes to already observe an average of 72\%,
and one in control of 30 most central nodes to be included in 90\% of payment paths.

In contrast, an adversary controlling randomly placed nodes may at best
observe an average of 5\% of payments.
Moreover, an adversary controlling the 26 \texttt{lnbig} nodes can observe
between 11\% and 25\% of payments, averaging at 15\%. 

Generally, the payments
with the highest amount result in the highest shares of compromised paths. 
This is most likely the case since more central nodes tend to optimize their
fee policies and are also well-connected capacity wise, \ie, are more likely
part of the few paths that can route higher-amount payments. However, one
exception to this rule can be observed in the case of \texttt{lnbig}, where
the 1 satoshi case yields the highest chance of path inclusion at
25\%. We assume this to be the case because of LNBIG's positioning in the
network and since their nodes feature a high amount of channels
with \texttt{base\_fee} set to 0, making them more likely to be chose by the
routing algorithm for low-amount payments.

\subsubsection{Adversarial Success}
\begin{figure*}
    \centering
    \input{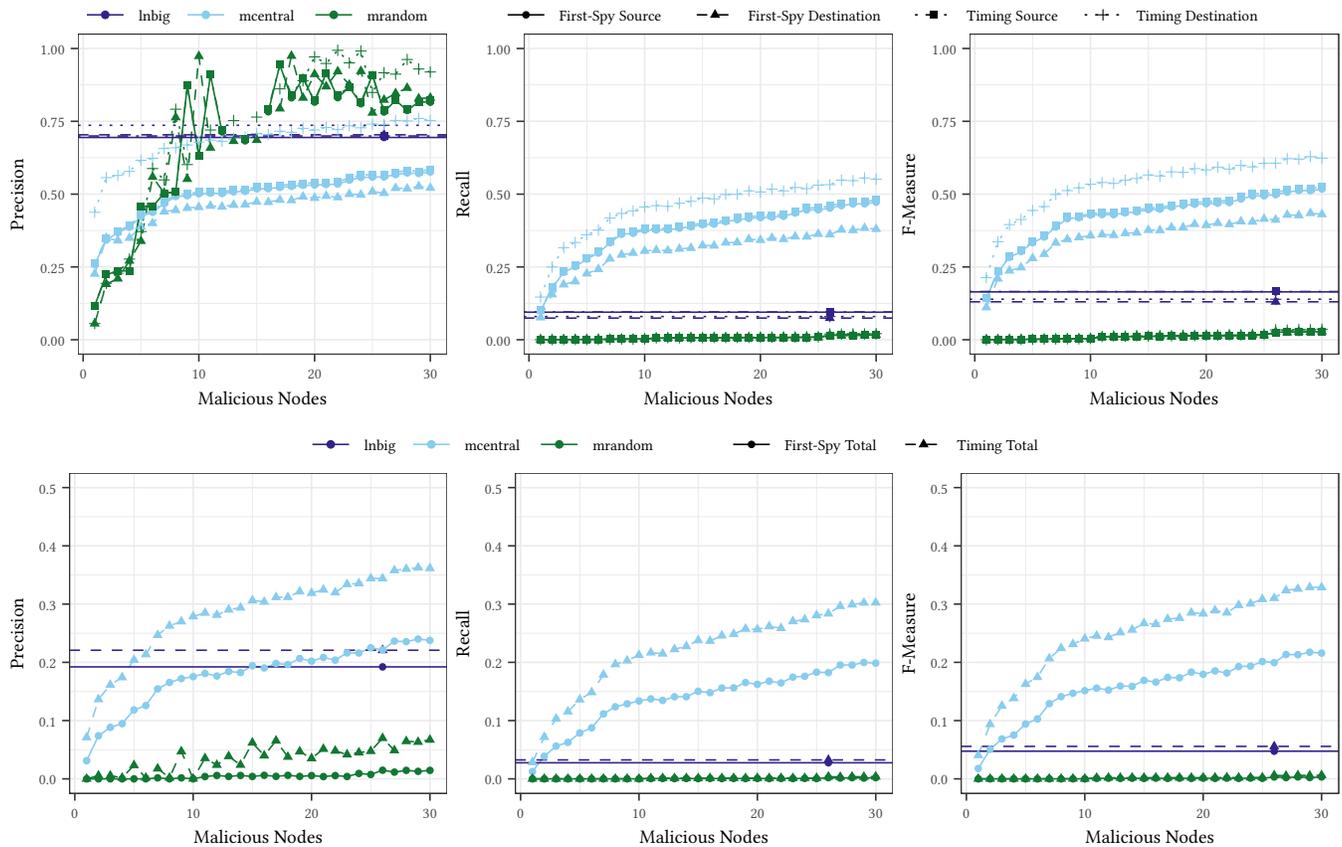}
    \caption{Adversarial success in dependence of number of malicious nodes.}
    \label{fig:plot_estimator_performance}
\end{figure*}

In the case of an adversary aiming to deanonymize payments, the performance with which
she can correctly guess the source or destination of a message is also a measure of (remaining) user privacy.
We therefore analyze how successful adversaries of different
magnitudes would be if they would run the proposed timing-based attacks by
applying the source estimator $\lnestimatorsymb_{s,\texttt{T}}$ and destination
estimator $\lnestimatorsymb_{t,\texttt{T}}$.
As a baseline for comparison, we also implemented and simulated the First-Spy
estimators $\lnestimatorsymb_{s,\texttt{FS}}$ and
$\lnestimatorsymb_{t,\texttt{FS}}$ that
respectively deem the predecessor of the first point of observation and
the successor of the last point of observation to be source and destination of
the observed payment.

The upper row of Figure~\ref{fig:plot_estimator_performance} shows the success
of the different estimators in dependence of the evaluated scenarios and
number of malicious nodes. As can be seen in
the top left plot, the precision with which the adversary estimates the
correct sources or destinations is generally correlated with the number of
controlled malicious nodes. In case of the \texttt{mcentral} scenario, the
precision of each estimator roughly follows a logarithmic growth function,
where the First-Spy destination estimator $\lnestimatorsymb_{t,\texttt{FS}}$
performs the worst ranging from $0.22$ for a single controlled node to $0.52$
for 30 malicious nodes. In contrast, the timing-based destination estimator
$\lnestimatorsymb_{t, \texttt{T}}$ yields the highest accuracy that ranges
from $0.45$ to $0.75$. Comparably, a potential adversary controlling the 26
\texttt{lnbig} nodes would be able to correctly identify senders or receivers with a
precision ranging from $0.69$ for $\lnestimatorsymb_{s,\texttt{FS}}$ to $0.73$
for $\lnestimatorsymb_{t, \texttt{T}}$. These high estimation results can be
attributed to the favorable positioning of LNBIG's nodes in the network
topology. In similar vein, it can be observed that
randomly placed malicious nodes, as in the \texttt{mrandom} scenrio, may
actually guess the correct senders and receivers with quite high accuracy, as
they cover the network graph more uniformly.

However, as shown in the top-middle plot of
Figure~\ref{fig:plot_estimator_performance}, \texttt{lnbig} and
\texttt{mrandom} nodes clearly do not perform as well in terms of recall.
While the share of correctly attributed payments barely reaches 2\% in the
best case ($\lnestimatorsymb_{t,\texttt{T}}$, $m = 30$), \texttt{lnbig} is
also only able to estimate the correct endpoints in 10\% of all payments at best.
Notably, the most central nodes have the highest recall, ranging between 7\%
($\lnestimatorsymb_{t,\texttt{FS}}$, $m = 1$) and 55\% of payments.
Of course, the recall is highly correlated with the share of observed payments
paths discussed in the previous subsection.

Therefore, in order to provide a unified measure that allows to analyze and compare the
overall accuracy of estimators, the top-right plot of
Figure~\ref{fig:plot_estimator_performance} shows the corresponding $F_1$-Measure.
In all cases, it shows that the First-Spy baseline is outperformed 
by the timing-based estimators, which reach up to an $F_1$ score of $0.62$ for
\texttt{mcentral}, $\lnestimatorsymb_{t,\texttt{T}}$, and $m=30$.
It is however noteworthy that while the timing-based source estimator always
performs better than its First-Spy counterpart, it doesn't do so by a
significant margin in some scenarios. This happens when the malicious nodes are 
placed close to the source node of the payment paths, which is generally the case for 
the high number of short payment paths and the more distributed node
positioning of the \texttt{lnbig} scenario in particular. Interestingly, we also found that the
weight-based routing algorithm (see Section~\ref{sec:attack:topoadv})
puts edges to more central (hence, in the \texttt{mcentral} case, more
malicious) nodes at the beginning of payment paths, which increases the success cases of the First-Spy
source estimator.

In order to give a overall comparison of timing-based attacks on privacy to
the First-Spy approach, we analyzed the number of payments that were
fully deanonymized, \ie, the number of payments for which the adversary was able to
correctly identify source \emph{and} destination. To this end, the bottom row
of Figure~\ref{fig:plot_estimator_performance} shows the precision, recall,
and $F_1$-measure with which the adversary could totally deanonymize payments
given the estimators $\lnestimatorsymb_{\texttt{FS}}$ and
$\lnestimatorsymb_{\texttt{T}}$. Of course, as this considers a subset of the
correct results of each individual estimator, all measures are lower.
However, the results generally follow the same behavior as just discussed.
Notably, the timing-based estimator outperforms the end-to-end deanonymization
performance of the First-Spy approach in every case of every scenario and in precision, recall, as well as
$F_1$-measure. It does so in particular in the \texttt{mcentral}
scenarios, in which it attack success is reliably higher than the baseline by
factor $1.5$. Thereby, our simulation results confirm the feasibility and
improved adversarial success of timing-based attacks on privacy in payment
channel networks.

\section{Discussion}\label{sec:discussion}
In the following, we discuss possible steps towards attack mitigation, the impact of upcoming changes
to the Lightning protocol, as well as avenues of future research.

\subsection{Possible Countermeasures}
The feasibility of timing attacks relies on the possibility to
build a reliable model of latencies, and on the adversary's capability of
observing and correlating of suitable
interactive multi-hop message exchanges, such as the current
\texttt{update\_add\_htlc}, \texttt{update\_fail\_htlc}, and \texttt{update\_fulfill\_htlc}
message payloads.

Therefore, in order to impair the retrieval of timing measurements, message
replies could be delayed for a random amount of time by the Lightning nodes,
along the lines of Bitcoin's transaction trickling
scheme~\cite{fanti2017deanon} or a timed mix network.
However, this would of course significantly delay payment processing and
therefore directly conflict with Lightning's goal of enabling quick payments.
It would furthermore counteract recent efforts to reduce end-to-end payment
latencies, such as Boomerang proposal~\cite{bagaria2019boomerang}.

Moreover, the adversary's capability of correlating payment observations could
be impaired, \eg, by introducing a payment scheme that does not leak 
identifying payment features, such as today's payment hash, such as anonymous
multi-hop locks~\cite{malavolta2019anonymous}. Note however, that even given
such a scheme, payment observations may still be correlated through metadata analysis,
as timing and payment amounts. Furthermore, as an individual node 
still needs to be able to match incoming and outgoing network messages, 
such decorrelation would only protect of re-identifying the same payment in the
network, \ie, mitigate full deanonymization. Very likely, the individual
source or destination estimators could still be applied.

\subsection{Impact of Protocol Changes}
The issue of \emph{payment path distinguishability} based on time-lock deltas was identified by the
developers of the Lightning protocol some time ago, which lead to the
introduction of so-called \emph{shadow routes} to the Lightning
standard~\cite{lnbolt07}. The idea behind shadow routes is to add a random padding to the
overall time-delta value of payments, so that the set of possible destinations would not
be identified by a remaining lock time of $0$. That said, different
implementations of Lightning handle the random padding differently, and, to
the best of our knowledge \texttt{LND} currently does not implement shadow
routes~\cite{lndshadowroutes} at all. In order to estimate what impact shadow routes
would have on the accuracy of timing-based destination estimators
$\lnestimatorsymb_{t, \texttt{T}}$, we re-evaluated the scenarios discussed in
Section~\ref{sec:eval:netsim} while disabling timelock-based anonymity set reduction.
Even though this corresponds to a worst-case estimation, we found the decrease
in precision and recall of the estimator to be only 2-3\%, indicating that
most of its performance is based on the timing-based maximum likelihood estimation.

The currently discussed proposal for \emph{Rendez-Vous
Routing}~\cite{lnrfcrendez} would allow for the creation of partial onion
messages that include the payloads for only a suffix of the payment path.
These partial onion messages could then be handed to an untrusted party, which
would be able to complete the payment path by supplying a suitable prefix to
the \emph{rendez-vous point}. This construction bears resemblance to Tor's
hidden services and would allow for receiver-anonymous payments, \ie, would
allow users to send payments whose location in the network they are not aware
of. While the implementation of this proposal would therefore generally
improve Lightning Network's privacy, it would likely not interfere
with the feasibility of timing attacks.

As sending large payments given the current channel capacities is often
unsuccessful, schemes allowing to split payments and route them over
different paths, such as the recently implemented \emph{multi-part payments},
have been discussed for some time in the Lightning community. As each
individual payment carries only part of the overall amount, they provide
increased value privacy, since the adversary is less likely to observe all
payments and cannot infer the actual transaction volume.
However, as this results in a higher number of closely correlated
payments, an adversary has a higher probability to observe such payments, whereby the
sender/receiver anonymity is decreased.

\subsection{Future Research}
Our analysis of timing attacks is based on the model of public Lightning
nodes, as introduced in Section~\ref{sec:model}. However, the Lightning
protocol also allows for the establishment
of \emph{hidden} payment channels that are only known to the
adjacent neighbors and are not broadcasted in the public peer-to-peer network.
As the estimators of course presuppose the knowledge of the underlying channel
graph to be able to return the candidate endpoint of maximum likelihood, they
are bound to fail in these circumstances. Therefore, applying methods from the
research area of \emph{topology inference}~\cite{neudecker2016timing,
daniel19mapz} in order to detect hidden channels would be an interesting
avenue for future research.

In our network simulations, we furthermore observed cases in which the
timing estimators wrongly identified the endpoints of unusually long payment
paths as the candidates with maximum likelihood. This is often the case when these
paths consist of many edges with small mean latencies, which then results in an
aggregated distribution that is closer to the measured time difference than
the correct candidate. Recently,
Kappos et al.~\cite{kappos2020empirical} proposed a
model for endpoint deanonymization based on a probability distribution over
payment path lengths. We think that integrating such an approach could help to
exclude such unusually long paths and hence further improve the results of
timing-based attacks on privacy.

\section{Related Work}\label{sec:relwork}
A large body of literature is concerned with the privacy of cryptocurrency
transactions. While initially most research mainly focussed on the privacy provisions
of the consensus layer~\cite{meiklejohn2013fistful, ron2012quantitative}, more
recent work discussed deanonymization attacks on the peer-to-peer
layer~\cite{koshy14p2p, biryukov2014deanonymisation}, what privacy properties
it can provide~\cite{fanti2017deanon}, and how it may be
improved~\cite{venkatakrishnan2017dandelion, fanti2018dandelion++}. 
Very recently, Tramer et al.~\cite{tramer2020remote} have shown that
timing-based side-channel information and traffic analysis may be used to
attack the privacy guarantees of Zcash and Monero cryptocurrency transactions.

Also, more and more work studies the possibility of attacks on the security and
privacy of second-layer solutions in general and payment channel networks in
particular. It has been shown that the Lightning
Network is vulnerable to channel exhaustion, node
isolation~\cite{rohrer19discharged}, as well as congestion
attacks~\cite{mizrahi2020congestion}.
Tochner et al.\ showed
that the routing algorithms employed by PCNs may be manipulated to in order to
facilitate inclusion of a malicious node in the payment path, thereby 
increasing the danger of denial-of-service attacks~\cite{tochner2019hijacking}.
Furthermore, recent entries have discussed the possibility of discovering the private channel
balances by probing~\cite{herrerajoancomarti2019difficulty,
vandam2019improvements} and analyzed how much privacy could be retained, if
noisy channel balances were to be made public~\cite{tang2019privacy}. 
A number of papers analyzed the graph-theoretic properties of the Lightning
Network graph and discussed possible consequences regarding decentralization
and routing~\cite{rohrer19discharged, seres2019topological,
lin2020secondpath}, as well as graph-based privacy
properties~\cite{martinazzi2020evolving}. 
Similarly, Beres et al.~\cite{beres2019cryptoeconomic}
and Tikhomirov et al.~\cite{tikhomirov2020quantitative}
empirically analyzed the privacy properties of PCNs with the assistance of model-based traffic simulation. 
While both of these entries discuss the possibility of deanonymization
attacks, they essentially apply a variant of the \emph{first-spy} estimator,
\ie, estimate immediate predecessors and successors to be senders/receivers.
Concurrently to our work, Kappos et al.~\cite{kappos2020empirical}
refine prior approaches of traffic simulation and introduce a probabilistic model
based on observed path lengths in order to estimate probable payment endpoints.
Moreover, while a recent entry by Nisslmüller et al.~\cite{nisslmueller2020toward}
mention the possibility of timing attacks, their
investigation remains in a preliminary state. 
To the best of our knowledge, our work is therefore the first to study the
impact and feasibility of timing attacks on privacy in payment channel networks in depth.

Orthogonally, a growing body of contributions aims to improve the privacy
guarantees of second-layer solutions.
While some designs allow to anonymously transact over private payment
hubs~\cite{heilman2017tumblebit, green2017bolt},
privacy-preserving routing mechanisms~\cite{roos2018settling,
mazumdar2020hushrelay} promise to enable anonymous transactions over multiple
intermediaries, \ie, payment channel networks.
Malavolta et al.\ proposed provably secure private payment
protocols~\cite{malavolta2017concurrency} and introduced a 
Lightning-compatible anonymous
locking mechanism based on ECDSA signatures that
allows for the decorrelation of payment paths~\cite{malavolta2019anonymous}. 
While they would not entirely mitigate the possibility of timing attacks, the
adoption of such anonymous multi-hop locks would force an adversary to take
additional error-prone measures for payment correlation, which would likely
result in reduced estimator precision.

\section{Conclusion}\label{sec:conclusion}
In this work, we studied potential timing attacks on privacy in payment
channel networks. We showed the feasibility of retrieving the required
time-differences through a proof-of-concept measurement node and evaluated the 
impact of this attack vector based on larger-scale network simulations.
As the results confirm the improved adversarial success, we conclude that
timing attacks on privacy may subvert Lightning's design goals of fast and
private cryptocurrency transactions.

\bibliographystyle{ACM-Reference-Format}
\bibliography{local,bib/anon,bib/blockchain,bib/p2p,bib/security,bib/dsi,bib/conferences-crossref}


\begin{thebibliography}{53}


\ifx \showCODEN    \undefined \def \showCODEN     #1{\unskip}     \fi
\ifx \showDOI      \undefined \def \showDOI       #1{#1}\fi
\ifx \showISBNx    \undefined \def \showISBNx     #1{\unskip}     \fi
\ifx \showISBNxiii \undefined \def \showISBNxiii  #1{\unskip}     \fi
\ifx \showISSN     \undefined \def \showISSN      #1{\unskip}     \fi
\ifx \showLCCN     \undefined \def \showLCCN      #1{\unskip}     \fi
\ifx \shownote     \undefined \def \shownote      #1{#1}          \fi
\ifx \showarticletitle \undefined \def \showarticletitle #1{#1}   \fi
\ifx \showURL      \undefined \def \showURL       {\relax}        \fi
\providecommand\bibfield[2]{#2}
\providecommand\bibinfo[2]{#2}
\providecommand\natexlab[1]{#1}
\providecommand\showeprint[2][]{arXiv:#2}

\bibitem[\protect\citeauthoryear{??}{ccs}{2017}]%
        {ccs17}
 \bibinfo{year}{2017}\natexlab{}.
\newblock \bibinfo{booktitle}{} (Dallas, TX, USA).
\newblock


\bibitem[\protect\citeauthoryear{Avarikioti, Heimbach, Wang, and
  Wattenhofer}{Avarikioti et~al\mbox{.}}{2019}]%
        {avarikioti2019ride}
\bibfield{author}{\bibinfo{person}{Zeta Avarikioti}, \bibinfo{person}{Lioba
  Heimbach}, \bibinfo{person}{Yuyi Wang}, {and} \bibinfo{person}{Roger
  Wattenhofer}.} \bibinfo{year}{2019}\natexlab{}.
\newblock \showarticletitle{Ride the Lightning: The Game Theory of Payment
  Channels}.
\newblock \bibinfo{journal}{\emph{CoRR}}  \bibinfo{volume}{abs/1912.04797}
  (\bibinfo{year}{2019}).
\newblock


\bibitem[\protect\citeauthoryear{Bagaria, Neu, and Tse}{Bagaria
  et~al\mbox{.}}{2019}]%
        {bagaria2019boomerang}
\bibfield{author}{\bibinfo{person}{Vivek Bagaria}, \bibinfo{person}{Joachim
  Neu}, {and} \bibinfo{person}{David Tse}.} \bibinfo{year}{2019}\natexlab{}.
\newblock \showarticletitle{Boomerang: Redundancy Improves Latency and
  Throughput in Payment Networks}.
\newblock \bibinfo{journal}{\emph{arXiv preprint arXiv:1910.01834}}
  (\bibinfo{year}{2019}).
\newblock


\bibitem[\protect\citeauthoryear{Bailey, Dittrich, Kenneally, and
  Maughan}{Bailey et~al\mbox{.}}{2012}]%
        {bailey2012menloreport}
\bibfield{author}{\bibinfo{person}{Michael Bailey}, \bibinfo{person}{David
  Dittrich}, \bibinfo{person}{Erin Kenneally}, {and} \bibinfo{person}{Douglas
  Maughan}.} \bibinfo{year}{2012}\natexlab{}.
\newblock \showarticletitle{The Menlo Report}.
\newblock \bibinfo{journal}{\emph{{IEEE} Secur. Priv.}} \bibinfo{volume}{10},
  \bibinfo{number}{2} (\bibinfo{year}{2012}), \bibinfo{pages}{71--75}.
\newblock


\bibitem[\protect\citeauthoryear{Bergamini, Crescenzi, D'Angelo, Meyerhenke,
  Severini, and Velaj}{Bergamini et~al\mbox{.}}{2018}]%
        {bergamini2018improving}
\bibfield{author}{\bibinfo{person}{Elisabetta Bergamini},
  \bibinfo{person}{Pierluigi Crescenzi}, \bibinfo{person}{Gianlorenzo
  D'Angelo}, \bibinfo{person}{Henning Meyerhenke}, \bibinfo{person}{Lorenzo
  Severini}, {and} \bibinfo{person}{Yllka Velaj}.}
  \bibinfo{year}{2018}\natexlab{}.
\newblock \showarticletitle{Improving the Betweenness Centrality of a Node by
  Adding Links}.
\newblock \bibinfo{journal}{\emph{{ACM} Journal of Experimental Algorithmics}}
  \bibinfo{volume}{23} (\bibinfo{year}{2018}).
\newblock


\bibitem[\protect\citeauthoryear{Biryukov, Khovratovich, and
  Pustogarov}{Biryukov et~al\mbox{.}}{[n.d.]}]%
        {biryukov2014deanonymisation}
\bibfield{author}{\bibinfo{person}{Alex Biryukov}, \bibinfo{person}{Dmitry
  Khovratovich}, {and} \bibinfo{person}{Ivan Pustogarov}.}
  \bibinfo{year}{[n.d.]}\natexlab{}.
\newblock \showarticletitle{Deanonymisation of Clients in Bitcoin P2P Network}.
  In \bibinfo{booktitle}{\emph{CCS~'14: Proceedings of the 21st ACM Conference
  on Computer and Communications Security}} (Scottsdale, AZ, USA, 2014-11).
  \bibinfo{pages}{15--29}.
\newblock


\bibitem[\protect\citeauthoryear{Bitnodes}{Bitnodes}{2020}]%
        {bitnodes}
\bibfield{author}{\bibinfo{person}{Bitnodes}.} \bibinfo{year}{2020}\natexlab{}.
\newblock \bibinfo{booktitle}{\emph{Homepage}}.
\newblock
\urldef\tempurl%
\url{https://bitnodes.io}
\showURL{%
Retrieved March 26, 2020 from \tempurl}


\bibitem[\protect\citeauthoryear{Brown}{Brown}{[n.d.]}]%
        {certicom2010eccparameters}
\bibfield{author}{\bibinfo{person}{Daniel R.~L. Brown}.}
  \bibinfo{year}{[n.d.]}\natexlab{}.
\newblock \bibinfo{title}{SEC 2: Recommended Elliptic Curve Domain Parameters}.
\newblock
\newblock
\urldef\tempurl%
\url{http://www.secg.org/sec2-v2.pdf}
\showURL{%
\tempurl}


\bibitem[\protect\citeauthoryear{Béres, Seres, and Benczúr}{Béres
  et~al\mbox{.}}{2019}]%
        {beres2019cryptoeconomic}
\bibfield{author}{\bibinfo{person}{Ferenc Béres},
  \bibinfo{person}{István~András Seres}, {and} \bibinfo{person}{András~A.
  Benczúr}.} \bibinfo{year}{2019}\natexlab{}.
\newblock \showarticletitle{A Cryptoeconomic Traffic Analysis of Bitcoins
  Lightning Network}.
\newblock \bibinfo{journal}{\emph{CoRR}}  \bibinfo{volume}{abs/1911.09432}
  (\bibinfo{year}{2019}).
\newblock
\showeprint[arxiv]{1911.09432}


\bibitem[\protect\citeauthoryear{{c-lightning Project}}{{c-lightning
  Project}}{2020}]%
        {clightning}
\bibfield{author}{\bibinfo{person}{{c-lightning Project}}.}
  \bibinfo{year}{2020}\natexlab{}.
\newblock \bibinfo{booktitle}{\emph{Github}}.
\newblock
\urldef\tempurl%
\url{https://github.com/ElementsProject/lightning}
\showURL{%
Retrieved May 8, 2020 from \tempurl}


\bibitem[\protect\citeauthoryear{Croman, Decker, Eyal, Gencer, Juels, Kosba,
  Miller, Saxena, Shi, Sirer, Song, and Wattenhofer}{Croman
  et~al\mbox{.}}{[n.d.]}]%
        {croman2016scalingblockchains}
\bibfield{author}{\bibinfo{person}{Kyle Croman}, \bibinfo{person}{Christian
  Decker}, \bibinfo{person}{Ittay Eyal}, \bibinfo{person}{Adem~Efe Gencer},
  \bibinfo{person}{Ari Juels}, \bibinfo{person}{Ahmed~E. Kosba},
  \bibinfo{person}{Andrew Miller}, \bibinfo{person}{Prateek Saxena},
  \bibinfo{person}{Elaine Shi}, \bibinfo{person}{Emin~Gün Sirer},
  \bibinfo{person}{Dawn Song}, {and} \bibinfo{person}{Roger Wattenhofer}.}
  \bibinfo{year}{[n.d.]}\natexlab{}.
\newblock \showarticletitle{On Scaling Decentralized Blockchains - A Position
  Paper}. In \bibinfo{booktitle}{\emph{BITCOIN~'16: Proceedings of the 3rd
  Workshop on Bitcoin Research}} (Christ Church, Barbados, 2016-02).
  \bibinfo{pages}{106--125}.
\newblock


\bibitem[\protect\citeauthoryear{Danezis and Goldberg}{Danezis and
  Goldberg}{2009}]%
        {danezis2009sphinx}
\bibfield{author}{\bibinfo{person}{George Danezis} {and} \bibinfo{person}{Ian
  Goldberg}.} \bibinfo{year}{2009}\natexlab{}.
\newblock \showarticletitle{Sphinx: {A} Compact and Provably Secure Mix
  Format}. In \bibinfo{booktitle}{\emph{SP~'09: Proceedings of the 30th IEEE
  Symposium on Security and Privacy}} (Oakland, CA, USA, 2009-05).
  \bibinfo{pages}{269--282}.
\newblock


\bibitem[\protect\citeauthoryear{Daniel, Rohrer, and Tschorsch}{Daniel
  et~al\mbox{.}}{[n.d.]}]%
        {daniel19mapz}
\bibfield{author}{\bibinfo{person}{Erik Daniel}, \bibinfo{person}{Elias
  Rohrer}, {and} \bibinfo{person}{Florian Tschorsch}.}
  \bibinfo{year}{[n.d.]}\natexlab{}.
\newblock \showarticletitle{Map-Z: Exposing the Zcash Network in Times of
  Transition}. In \bibinfo{booktitle}{\emph{LCN~'19: Proceedings of the 44th
  IEEE International Conference on Local Computer Networks}} (Osnabrück,
  Germany, 2019-10).
\newblock


\bibitem[\protect\citeauthoryear{Decker}{Decker}{[n.d.]}]%
        {lnrfcrendez}
\bibfield{author}{\bibinfo{person}{Christian Decker}.}
  \bibinfo{year}{[n.d.]}\natexlab{}.
\newblock \bibinfo{title}{Rendez-Vous Routing Proposal}.
\newblock
\newblock
\urldef\tempurl%
\url{https://github.com/lightningnetwork/lightning-rfc/blob/rendez-vous/proposals/0001-rendez-vous.md}
\showURL{%
\tempurl}


\bibitem[\protect\citeauthoryear{Dijkstra}{Dijkstra}{1959}]%
        {dijkstra1959note}
\bibfield{author}{\bibinfo{person}{Edsger~W. Dijkstra}.}
  \bibinfo{year}{1959}\natexlab{}.
\newblock \showarticletitle{A note on two problems in connexion with graphs}.
\newblock \bibinfo{journal}{\emph{Numer. Math.}}  \bibinfo{volume}{1}
  (\bibinfo{year}{1959}), \bibinfo{pages}{269--271}.
\newblock


\bibitem[\protect\citeauthoryear{Ersoy, Roos, and Erkin}{Ersoy
  et~al\mbox{.}}{2019}]%
        {ersoy2019profit}
\bibfield{author}{\bibinfo{person}{Oguzhan Ersoy}, \bibinfo{person}{Stefanie
  Roos}, {and} \bibinfo{person}{Zekeriya Erkin}.}
  \bibinfo{year}{2019}\natexlab{}.
\newblock \showarticletitle{How to profit from payments channels}.
\newblock \bibinfo{journal}{\emph{CoRR}}  \bibinfo{volume}{abs/1911.08803}
  (\bibinfo{year}{2019}).
\newblock


\bibitem[\protect\citeauthoryear{Fanti, Venkatakrishnan, Bakshi, Denby,
  Bhargava, Miller, and Viswanath}{Fanti et~al\mbox{.}}{[n.d.]}]%
        {fanti2018dandelion++}
\bibfield{author}{\bibinfo{person}{Giulia~C. Fanti},
  \bibinfo{person}{Shaileshh~Bojja Venkatakrishnan}, \bibinfo{person}{Surya
  Bakshi}, \bibinfo{person}{Bradley Denby}, \bibinfo{person}{Shruti Bhargava},
  \bibinfo{person}{Andrew Miller}, {and} \bibinfo{person}{Pramod Viswanath}.}
  \bibinfo{year}{[n.d.]}\natexlab{}.
\newblock \showarticletitle{Dandelion++: Lightweight Cryptocurrency Networking
  with Formal Anonymity Guarantees}.
\newblock  \bibinfo{volume}{2}, \bibinfo{number}{2} (\bibinfo{year}{[n.\,d.]}),
  \bibinfo{pages}{29:1--29:35}.
\newblock


\bibitem[\protect\citeauthoryear{Fanti and Viswanath}{Fanti and
  Viswanath}{[n.d.]}]%
        {fanti2017deanon}
\bibfield{author}{\bibinfo{person}{Giulia~C. Fanti} {and}
  \bibinfo{person}{Pramod Viswanath}.} \bibinfo{year}{[n.d.]}\natexlab{}.
\newblock \showarticletitle{Deanonymization in the Bitcoin {P2P} Network}. In
  \bibinfo{booktitle}{\emph{NIPS~'17: Proceedings of 30th Annual Conference on
  Neural Information Processing Systems}} (Long Beach, CA, {USA}, 2017-12).
\newblock


\bibitem[\protect\citeauthoryear{Green and Miers}{Green and Miers}{2017}]%
        {green2017bolt}
\bibfield{author}{\bibinfo{person}{Matthew Green} {and} \bibinfo{person}{Ian
  Miers}.} \bibinfo{year}{2017}\natexlab{}.
\newblock \showarticletitle{Bolt: Anonymous Payment Channels for Decentralized
  Currencies}, See \citeN{ccs17}, \bibinfo{pages}{473--489}.
\newblock


\bibitem[\protect\citeauthoryear{Heilman, Alshenibr, Baldimtsi, Scafuro, and
  Goldberg}{Heilman et~al\mbox{.}}{2017}]%
        {heilman2017tumblebit}
\bibfield{author}{\bibinfo{person}{Ethan Heilman}, \bibinfo{person}{Leen
  Alshenibr}, \bibinfo{person}{Foteini Baldimtsi}, \bibinfo{person}{Alessandra
  Scafuro}, {and} \bibinfo{person}{Sharon Goldberg}.}
  \bibinfo{year}{2017}\natexlab{}.
\newblock \showarticletitle{TumbleBit: An Untrusted Bitcoin-Compatible
  Anonymous Payment Hub}. In \bibinfo{booktitle}{\emph{NDSS~'17: Proceedings of
  the 24th Annual Network and Distributed System Security Symposium}} (San
  Diego, California, USA).
\newblock


\bibitem[\protect\citeauthoryear{Herrera-Joancomartí, Navarro-Arribas,
  Pedrosa, Pérez-Solà, and García-Alfaro}{Herrera-Joancomartí
  et~al\mbox{.}}{2019}]%
        {herrerajoancomarti2019difficulty}
\bibfield{author}{\bibinfo{person}{Jordi Herrera-Joancomartí},
  \bibinfo{person}{Guillermo Navarro-Arribas},
  \bibinfo{person}{Alejandro~Ranchal Pedrosa}, \bibinfo{person}{Cristina
  Pérez-Solà}, {and} \bibinfo{person}{Joaquín García-Alfaro}.}
  \bibinfo{year}{2019}\natexlab{}.
\newblock \showarticletitle{On the Difficulty of Hiding the Balance of
  Lightning Network Channels}. In \bibinfo{booktitle}{\emph{AsiaCCS~'19:
  Proceedings of the 2019 {ACM} Asia Conference on Computer and Communications
  Security}} (Auckland, New Zealand). \bibinfo{pages}{602--612}.
\newblock


\bibitem[\protect\citeauthoryear{Johnson, Wacek, Jansen, Sherr, and
  Syverson}{Johnson et~al\mbox{.}}{[n.d.]}]%
        {johnson2013usersrouted}
\bibfield{author}{\bibinfo{person}{Aaron Johnson}, \bibinfo{person}{Chris
  Wacek}, \bibinfo{person}{Rob Jansen}, \bibinfo{person}{Micah Sherr}, {and}
  \bibinfo{person}{Paul Syverson}.} \bibinfo{year}{[n.d.]}\natexlab{}.
\newblock \showarticletitle{Users Get Routed: Traffic Correlation on {Tor} by
  Realistic Adversaries}. In \bibinfo{booktitle}{\emph{CCS~'13: Proceedings of
  the 20th ACM Conference on Computer and Communications Security}} (Berlin,
  Germany, 2013-10). \bibinfo{pages}{337--348}.
\newblock


\bibitem[\protect\citeauthoryear{Kappos, Yousaf, Piotrowska, Kanjalkar,
  Delgado-Segura, Miller, and Meiklejohn}{Kappos et~al\mbox{.}}{2020}]%
        {kappos2020empirical}
\bibfield{author}{\bibinfo{person}{George Kappos}, \bibinfo{person}{Haaroon
  Yousaf}, \bibinfo{person}{Ania Piotrowska}, \bibinfo{person}{Sanket
  Kanjalkar}, \bibinfo{person}{Sergi Delgado-Segura}, \bibinfo{person}{Andrew
  Miller}, {and} \bibinfo{person}{Sarah Meiklejohn}.}
  \bibinfo{year}{2020}\natexlab{}.
\newblock \showarticletitle{An Empirical Analysis of Privacy in the Lightning
  Network}.
\newblock \bibinfo{journal}{\emph{arXiv preprint arXiv:2003.12470}}
  (\bibinfo{year}{2020}).
\newblock


\bibitem[\protect\citeauthoryear{Koshy, Koshy, and McDaniel}{Koshy
  et~al\mbox{.}}{[n.d.]}]%
        {koshy14p2p}
\bibfield{author}{\bibinfo{person}{Philip Koshy}, \bibinfo{person}{Diana
  Koshy}, {and} \bibinfo{person}{Patrick McDaniel}.}
  \bibinfo{year}{[n.d.]}\natexlab{}.
\newblock \showarticletitle{An Analysis of Anonymity in Bitcoin Using P2P
  Network Traffic}. In \bibinfo{booktitle}{\emph{FC~'14: Proceedings of the
  18th International Conference on Financial Cryptography and Data Security}}
  (Barbados, 2014-03). \bibinfo{pages}{469--485}.
\newblock


\bibitem[\protect\citeauthoryear{Lin, Primicerio, Squartini, Decker, and
  Tessone}{Lin et~al\mbox{.}}{2020}]%
        {lin2020secondpath}
\bibfield{author}{\bibinfo{person}{Jian{-}Hong Lin}, \bibinfo{person}{Kevin
  Primicerio}, \bibinfo{person}{Tiziano Squartini}, \bibinfo{person}{Christian
  Decker}, {and} \bibinfo{person}{Claudio~J. Tessone}.}
  \bibinfo{year}{2020}\natexlab{}.
\newblock \showarticletitle{Lightning Network: a second path towards
  centralisation of the Bitcoin economy}.
\newblock \bibinfo{journal}{\emph{CoRR}}  \bibinfo{volume}{abs/2002.02819}
  (\bibinfo{year}{2020}).
\newblock
\showeprint[arxiv]{2002.02819}


\bibitem[\protect\citeauthoryear{{LND}}{{LND}}{2020a}]%
        {lndsecondchancecommit}
\bibfield{author}{\bibinfo{person}{{LND}}.} \bibinfo{year}{2020}\natexlab{a}.
\newblock \bibinfo{booktitle}{\emph{Github Commit: Move Second Chance Logic}}.
\newblock
\urldef\tempurl%
\url{https://github.com/lightningnetwork/lnd/commit/dc13da5abbfa429273b516abd566f6c6fa5bb200}
\showURL{%
Retrieved May 8, 2020 from \tempurl}


\bibitem[\protect\citeauthoryear{{LND}}{{LND}}{2020b}]%
        {lndpolicyfailure}
\bibfield{author}{\bibinfo{person}{{LND}}.} \bibinfo{year}{2020}\natexlab{b}.
\newblock \bibinfo{booktitle}{\emph{Github: Policy Failure Logic}}.
\newblock
\urldef\tempurl%
\url{https://github.com/lightningnetwork/lnd/blob/1354a461701b9396f0b4a35b01d308c5fcc0dbd2/routing/result_interpretation.go#L343}
\showURL{%
Retrieved May 8, 2020 from \tempurl}


\bibitem[\protect\citeauthoryear{{LND}}{{LND}}{[n.d.]}]%
        {lndshadowroutes}
\bibfield{author}{\bibinfo{person}{Lightning Network~Daemon {LND}}.}
  \bibinfo{year}{[n.d.]}\natexlab{}.
\newblock \bibinfo{title}{Shadow Route Github Issue}.
\newblock
\newblock
\urldef\tempurl%
\url{https://github.com/lightningnetwork/lnd/issues/1222}
\showURL{%
\tempurl}


\bibitem[\protect\citeauthoryear{Malavolta, Moreno{-}Sanchez, Kate, Maffei, and
  Ravi}{Malavolta et~al\mbox{.}}{2017}]%
        {malavolta2017concurrency}
\bibfield{author}{\bibinfo{person}{Giulio Malavolta}, \bibinfo{person}{Pedro
  Moreno{-}Sanchez}, \bibinfo{person}{Aniket Kate}, \bibinfo{person}{Matteo
  Maffei}, {and} \bibinfo{person}{Srivatsan Ravi}.}
  \bibinfo{year}{2017}\natexlab{}.
\newblock \showarticletitle{Concurrency and Privacy with Payment-Channel
  Networks}, See \citeN{ccs17}, \bibinfo{pages}{455--471}.
\newblock


\bibitem[\protect\citeauthoryear{Malavolta, Moreno{-}Sanchez, Schneidewind,
  Kate, and Maffei}{Malavolta et~al\mbox{.}}{2019}]%
        {malavolta2019anonymous}
\bibfield{author}{\bibinfo{person}{Giulio Malavolta}, \bibinfo{person}{Pedro
  Moreno{-}Sanchez}, \bibinfo{person}{Clara Schneidewind},
  \bibinfo{person}{Aniket Kate}, {and} \bibinfo{person}{Matteo Maffei}.}
  \bibinfo{year}{2019}\natexlab{}.
\newblock \showarticletitle{Anonymous Multi-Hop Locks for Blockchain
  Scalability and Interoperability}. In \bibinfo{booktitle}{\emph{NDSS~'19:
  Prooceedings of the 26th Annual Network and Distributed System Security
  Symposium}} (San Diego, California, USA).
\newblock


\bibitem[\protect\citeauthoryear{Martinazzi and Flori}{Martinazzi and
  Flori}{2020}]%
        {martinazzi2020evolving}
\bibfield{author}{\bibinfo{person}{Stefano Martinazzi} {and}
  \bibinfo{person}{Andrea Flori}.} \bibinfo{year}{2020}\natexlab{}.
\newblock \showarticletitle{The evolving topology of the Lightning Network:
  Centralization, efficiency, robustness, synchronization, and anonymity}.
\newblock \bibinfo{journal}{\emph{PloS one}} \bibinfo{volume}{15},
  \bibinfo{number}{1} (\bibinfo{year}{2020}), \bibinfo{pages}{e0225966}.
\newblock


\bibitem[\protect\citeauthoryear{MaxMind}{MaxMind}{2020}]%
        {geolite}
\bibfield{author}{\bibinfo{person}{Inc. MaxMind}.}
  \bibinfo{year}{2020}\natexlab{}.
\newblock \bibinfo{booktitle}{\emph{GeoIP GeoLite2 database}}.
\newblock
\urldef\tempurl%
\url{https://dev.maxmind.com/geoip/geoip2/geolite2/}
\showURL{%
Retrieved March 26, 2020 from \tempurl}


\bibitem[\protect\citeauthoryear{Mazumdar, Ruj, Singh, and Pal}{Mazumdar
  et~al\mbox{.}}{2020}]%
        {mazumdar2020hushrelay}
\bibfield{author}{\bibinfo{person}{Subhra Mazumdar}, \bibinfo{person}{Sushmita
  Ruj}, \bibinfo{person}{Ram~Govind Singh}, {and} \bibinfo{person}{Arindam
  Pal}.} \bibinfo{year}{2020}\natexlab{}.
\newblock \showarticletitle{HushRelay: {A} Privacy-Preserving, Efficient, and
  Scalable Routing Algorithm for Off-Chain Payments}.
\newblock \bibinfo{journal}{\emph{CoRR}}  \bibinfo{volume}{abs/2002.05071}
  (\bibinfo{year}{2020}).
\newblock
\showeprint[arxiv]{2002.05071}


\bibitem[\protect\citeauthoryear{Meiklejohn, Pomarole, Jordan, Levchenko,
  McCoy, Voelker, and Savage}{Meiklejohn et~al\mbox{.}}{[n.d.]}]%
        {meiklejohn2013fistful}
\bibfield{author}{\bibinfo{person}{Sarah Meiklejohn}, \bibinfo{person}{Marjori
  Pomarole}, \bibinfo{person}{Grant Jordan}, \bibinfo{person}{Kirill
  Levchenko}, \bibinfo{person}{Damon McCoy}, \bibinfo{person}{Geoffrey~M
  Voelker}, {and} \bibinfo{person}{Stefan Savage}.}
  \bibinfo{year}{[n.d.]}\natexlab{}.
\newblock \showarticletitle{A fistful of bitcoins: characterizing payments
  among men with no names}. In \bibinfo{booktitle}{\emph{IMC~'13: Proceedings
  of the 13th ACM SIGCOMM Conference on Internet Measurement}} (Barcelona,
  Spain, 2013-10). \bibinfo{pages}{127--140}.
\newblock


\bibitem[\protect\citeauthoryear{Mizrahi and Zohar}{Mizrahi and Zohar}{2020}]%
        {mizrahi2020congestion}
\bibfield{author}{\bibinfo{person}{Ayelet Mizrahi} {and} \bibinfo{person}{Aviv
  Zohar}.} \bibinfo{year}{2020}\natexlab{}.
\newblock \showarticletitle{Congestion Attacks in Payment Channel Networks}.
\newblock \bibinfo{journal}{\emph{CoRR}}  \bibinfo{volume}{abs/2002.06564}
  (\bibinfo{year}{2020}).
\newblock
\showeprint[arxiv]{2002.06564}


\bibitem[\protect\citeauthoryear{Nakamoto}{Nakamoto}{[n.d.]}]%
        {nakamoto2008bitcoin}
\bibfield{author}{\bibinfo{person}{Satoshi Nakamoto}.}
  \bibinfo{year}{[n.d.]}\natexlab{}.
\newblock \bibinfo{title}{Bitcoin: A peer-to-peer electronic cash system}.
\newblock
\newblock


\bibitem[\protect\citeauthoryear{Network}{Network}{[n.d.]a}]%
        {lnbolt04}
\bibfield{author}{\bibinfo{person}{Lightning Network}.}
  \bibinfo{year}{[n.d.]}\natexlab{a}.
\newblock \bibinfo{title}{{BOLT} \#4: Onion Routing Protocol}.
\newblock
\newblock
\urldef\tempurl%
\url{https://github.com/lightningnetwork/lightning-rfc/blob/master/04-onion-routing.md}
\showURL{%
\tempurl}


\bibitem[\protect\citeauthoryear{Network}{Network}{[n.d.]b}]%
        {lnbolt07}
\bibfield{author}{\bibinfo{person}{Lightning Network}.}
  \bibinfo{year}{[n.d.]}\natexlab{b}.
\newblock \bibinfo{title}{{BOLT} \#7: {P2P} Node and Channel Discovery}.
\newblock
\newblock
\urldef\tempurl%
\url{https://github.com/lightningnetwork/lightning-rfc/blob/master/07-routing-gossip.md}
\showURL{%
\tempurl}


\bibitem[\protect\citeauthoryear{Network}{Network}{[n.d.]c}]%
        {lnrfc}
\bibfield{author}{\bibinfo{person}{Lightning Network}.}
  \bibinfo{year}{[n.d.]}\natexlab{c}.
\newblock \bibinfo{title}{{BOLT} In-Progress Specifications}.
\newblock
\newblock
\urldef\tempurl%
\url{https://github.com/lightningnetwork/lightning-rfc}
\showURL{%
\tempurl}


\bibitem[\protect\citeauthoryear{Neudecker, Andelfinger, and
  Hartenstein}{Neudecker et~al\mbox{.}}{[n.d.]}]%
        {neudecker2016timing}
\bibfield{author}{\bibinfo{person}{Till Neudecker}, \bibinfo{person}{Philipp
  Andelfinger}, {and} \bibinfo{person}{Hannes Hartenstein}.}
  \bibinfo{year}{[n.d.]}\natexlab{}.
\newblock \showarticletitle{Timing Analysis for Inferring the Topology of the
  Bitcoin Peer-to-Peer Network}. In \bibinfo{booktitle}{\emph{UIC~'16:
  Proceedings of the 2016 International Conference on Ubiquitous Intelligence
  {\&} Computing}} (Toulouse, France, 2016-07).
\newblock


\bibitem[\protect\citeauthoryear{Nisslmueller, Foerster, Schmid, and
  Decker}{Nisslmueller et~al\mbox{.}}{2020}]%
        {nisslmueller2020toward}
\bibfield{author}{\bibinfo{person}{Utz Nisslmueller},
  \bibinfo{person}{Klaus-Tycho Foerster}, \bibinfo{person}{Stefan Schmid},
  {and} \bibinfo{person}{Christian Decker}.} \bibinfo{year}{2020}\natexlab{}.
\newblock \showarticletitle{Toward Active and Passive Confidentiality Attacks
  On Cryptocurrency Off-Chain Networks}. In
  \bibinfo{booktitle}{\emph{ICISSP~'20: Proceedings of the 6th International
  Conference on Information Systems Security and Privacy}} (Valetta, Malta).
\newblock


\bibitem[\protect\citeauthoryear{Perrin}{Perrin}{[n.d.]}]%
        {perrin2018noise}
\bibfield{author}{\bibinfo{person}{Trevor Perrin}.}
  \bibinfo{year}{[n.d.]}\natexlab{}.
\newblock \bibinfo{title}{The Noise Protocol Framework}.
\newblock
\newblock
\urldef\tempurl%
\url{https://noiseprotocol.org/noise.pdf}
\showURL{%
\tempurl}


\bibitem[\protect\citeauthoryear{Poon and Dryja}{Poon and Dryja}{[n.d.]}]%
        {poon2015bitcoin}
\bibfield{author}{\bibinfo{person}{Joseph Poon} {and} \bibinfo{person}{Thaddeus
  Dryja}.} \bibinfo{year}{[n.d.]}\natexlab{}.
\newblock \showarticletitle{The bitcoin lightning network: Scalable off-chain
  instant payments}.
\newblock  (\bibinfo{year}{[n.\,d.]}).
\newblock


\bibitem[\protect\citeauthoryear{Rohrer, Malliaris, and Tschorsch}{Rohrer
  et~al\mbox{.}}{[n.d.]}]%
        {rohrer19discharged}
\bibfield{author}{\bibinfo{person}{Elias Rohrer}, \bibinfo{person}{Julian
  Malliaris}, {and} \bibinfo{person}{Florian Tschorsch}.}
  \bibinfo{year}{[n.d.]}\natexlab{}.
\newblock \showarticletitle{Discharged Payment Channels: Quantifying the
  Lightning Network's Resilience to Topology-Based Attacks}. In
  \bibinfo{booktitle}{\emph{S{\&}B '19: Proceedings of IEEE Security {\&}
  Privacy on the Blockchain}} (2019-06).
\newblock


\bibitem[\protect\citeauthoryear{Ron and Shamir}{Ron and Shamir}{[n.d.]}]%
        {ron2012quantitative}
\bibfield{author}{\bibinfo{person}{Dorit Ron} {and} \bibinfo{person}{Adi
  Shamir}.} \bibinfo{year}{[n.d.]}\natexlab{}.
\newblock \showarticletitle{Quantitative Analysis of the Full Bitcoin
  Transaction Graph}. In \bibinfo{booktitle}{\emph{FC~'13: Proceedings of the
  17th International Conference on Financial Cryptography and Data Security}}
  (Okinawa, Japan, 2013-04). \bibinfo{pages}{6--24}.
\newblock


\bibitem[\protect\citeauthoryear{Roos, Moreno-Sanchez, Kate, and Goldberg}{Roos
  et~al\mbox{.}}{[n.d.]}]%
        {roos2018settling}
\bibfield{author}{\bibinfo{person}{Stefanie Roos}, \bibinfo{person}{Pedro
  Moreno-Sanchez}, \bibinfo{person}{Aniket Kate}, {and} \bibinfo{person}{Ian
  Goldberg}.} \bibinfo{year}{[n.d.]}\natexlab{}.
\newblock \showarticletitle{Settling Payments Fast and Private: Efficient
  Decentralized Routing for Path-Based Transactions}.
\newblock


\bibitem[\protect\citeauthoryear{Seres, Guly{\'{a}}s, Nagy, and Burcsi}{Seres
  et~al\mbox{.}}{2019}]%
        {seres2019topological}
\bibfield{author}{\bibinfo{person}{Istv{\'{a}}n~Andr{\'{a}}s Seres},
  \bibinfo{person}{L{\'{a}}szl{\'{o}} Guly{\'{a}}s},
  \bibinfo{person}{D{\'{a}}niel~A. Nagy}, {and} \bibinfo{person}{P{\'{e}}ter
  Burcsi}.} \bibinfo{year}{2019}\natexlab{}.
\newblock \showarticletitle{Topological Analysis of Bitcoin's Lightning
  Network}.
\newblock \bibinfo{journal}{\emph{CoRR}}  \bibinfo{volume}{abs/1901.04972}
  (\bibinfo{year}{2019}).
\newblock
\showeprint[arxiv]{1901.04972}


\bibitem[\protect\citeauthoryear{Tang, Wang, Fanti, and Oh}{Tang
  et~al\mbox{.}}{2019}]%
        {tang2019privacy}
\bibfield{author}{\bibinfo{person}{Weizhao Tang}, \bibinfo{person}{Weina Wang},
  \bibinfo{person}{Giulia~C. Fanti}, {and} \bibinfo{person}{Sewoong Oh}.}
  \bibinfo{year}{2019}\natexlab{}.
\newblock \showarticletitle{Privacy-Utility Tradeoffs in Routing Cryptocurrency
  over Payment Channel Networks}.
\newblock \bibinfo{journal}{\emph{CoRR}}  \bibinfo{volume}{abs/1909.02717}
  (\bibinfo{year}{2019}).
\newblock
\showeprint[arxiv]{1909.02717}


\bibitem[\protect\citeauthoryear{Tikhomirov, Moreno-Sanchez, and
  Maffei}{Tikhomirov et~al\mbox{.}}{[n.d.]}]%
        {tikhomirov2020quantitative}
\bibfield{author}{\bibinfo{person}{Sergei Tikhomirov}, \bibinfo{person}{Pedro
  Moreno-Sanchez}, {and} \bibinfo{person}{Matteo Maffei}.}
  \bibinfo{year}{[n.d.]}\natexlab{}.
\newblock \showarticletitle{A Quantitative Analysis of Security, Anonymity and
  Scalability for the Lightning Network}.
\newblock  (\bibinfo{year}{[n.\,d.]}).
\newblock


\bibitem[\protect\citeauthoryear{Tochner, Schmid, and Zohar}{Tochner
  et~al\mbox{.}}{2019}]%
        {tochner2019hijacking}
\bibfield{author}{\bibinfo{person}{Saar Tochner}, \bibinfo{person}{Stefan
  Schmid}, {and} \bibinfo{person}{Aviv Zohar}.}
  \bibinfo{year}{2019}\natexlab{}.
\newblock \showarticletitle{Hijacking Routes in Payment Channel Networks: {A}
  Predictability Tradeoff}.
\newblock \bibinfo{journal}{\emph{CoRR}}  \bibinfo{volume}{abs/1909.06890}
  (\bibinfo{year}{2019}).
\newblock
\showeprint[arxiv]{1909.06890}


\bibitem[\protect\citeauthoryear{Tram{\`{e}}r, Boneh, and
  Paterson}{Tram{\`{e}}r et~al\mbox{.}}{2020}]%
        {tramer2020remote}
\bibfield{author}{\bibinfo{person}{Florian Tram{\`{e}}r}, \bibinfo{person}{Dan
  Boneh}, {and} \bibinfo{person}{Kenneth~G. Paterson}.}
  \bibinfo{year}{2020}\natexlab{}.
\newblock \showarticletitle{Remote Side-Channel Attacks on Anonymous
  Transactions}.
\newblock \bibinfo{journal}{\emph{{IACR} Cryptology ePrint Archive}}
  \bibinfo{volume}{2020} (\bibinfo{year}{2020}), \bibinfo{pages}{220}.
\newblock


\bibitem[\protect\citeauthoryear{van Dam, Kadir, Nohuddin, and Zaman}{van Dam
  et~al\mbox{.}}{2019}]%
        {vandam2019improvements}
\bibfield{author}{\bibinfo{person}{Gijs van Dam}, \bibinfo{person}{Rabiah~Abdul
  Kadir}, \bibinfo{person}{Puteri N.~E. Nohuddin}, {and}
  \bibinfo{person}{Halimah~Badioze Zaman}.} \bibinfo{year}{2019}\natexlab{}.
\newblock \showarticletitle{Improvements of the Balance Discovery Attack on
  Lightning Network Payment Channels}.
\newblock \bibinfo{journal}{\emph{{IACR} Cryptology ePrint Archive}}
  \bibinfo{volume}{2019} (\bibinfo{year}{2019}), \bibinfo{pages}{1385}.
\newblock


\bibitem[\protect\citeauthoryear{Venkatakrishnan, Fanti, and
  Viswanath}{Venkatakrishnan et~al\mbox{.}}{[n.d.]}]%
        {venkatakrishnan2017dandelion}
\bibfield{author}{\bibinfo{person}{Shaileshh~Bojja Venkatakrishnan},
  \bibinfo{person}{Giulia~C. Fanti}, {and} \bibinfo{person}{Pramod Viswanath}.}
  \bibinfo{year}{[n.d.]}\natexlab{}.
\newblock \showarticletitle{Dandelion: Redesigning the Bitcoin Network for
  Anonymity}.
\newblock  (\bibinfo{year}{[n.\,d.]}).
\newblock


\end{thebibliography}
\appendix

\section{Lightning's Peer-to-Peer Network}\label{app:lnp2p}
In order to acquire a reliable model for inter-peer
connections, we in the following examine Lightning's public peer-to-peer network. To this end,
we acquired a snapshot\footnote{\url{https://gitlab.tu-berlin.de/rohrer/discharged-pc-data/blob/master/snapshots/lngraph_2020_03_26__00_00.json.zst}} of the network graph taken on March
26, 2020 00:00 UTC, extracted the $2679$ public IPv4 addresses, and categorized them in regional clusters based on
GeoLite2~\cite{geolite} geographic location database. As shown in
Figure~\ref{fig:plot_peers}, the peer-to-peer network spans seven regions of
the globe: Europe (EU), North America (NA), Asia (AS), Oceania (OC), South
America (SA), China (CN), and Africa (AF). The data however also shows that the network is
currently clearly dominated by the EU and NA regional clusters, which is in
accordance with the regional distribution of Bitcoin's peer-to-peer network~\cite{bitnodes}.

\begin{figure}[]
    \centering
    \begin{tikzpicture}[x=1pt,y=1pt]
\definecolor{fillColor}{RGB}{255,255,255}
\path[use as bounding box,fill=fillColor,fill opacity=0.00] (0,0) rectangle (238.49,108.41);
\begin{scope}
\path[clip] (  0.00,  0.00) rectangle (238.49,108.41);
\definecolor{drawColor}{RGB}{255,255,255}
\definecolor{fillColor}{RGB}{255,255,255}

\path[draw=drawColor,line width= 0.6pt,line join=round,line cap=round,fill=fillColor] (  0.00,  0.00) rectangle (238.49,108.41);
\end{scope}
\begin{scope}
\path[clip] ( 30.41, 23.26) rectangle (235.65,105.56);
\definecolor{fillColor}{RGB}{255,255,255}

\path[fill=fillColor] ( 30.41, 23.26) rectangle (235.65,105.56);
\definecolor{drawColor}{gray}{0.92}

\path[draw=drawColor,line width= 0.3pt,line join=round] ( 30.41, 40.90) --
	(235.65, 40.90);

\path[draw=drawColor,line width= 0.3pt,line join=round] ( 30.41, 68.74) --
	(235.65, 68.74);

\path[draw=drawColor,line width= 0.3pt,line join=round] ( 30.41, 96.58) --
	(235.65, 96.58);

\path[draw=drawColor,line width= 0.6pt,line join=round] ( 30.41, 26.98) --
	(235.65, 26.98);

\path[draw=drawColor,line width= 0.6pt,line join=round] ( 30.41, 54.82) --
	(235.65, 54.82);

\path[draw=drawColor,line width= 0.6pt,line join=round] ( 30.41, 82.66) --
	(235.65, 82.66);

\path[draw=drawColor,line width= 0.6pt,line join=round] ( 47.51, 23.26) --
	( 47.51,105.56);

\path[draw=drawColor,line width= 0.6pt,line join=round] ( 76.02, 23.26) --
	( 76.02,105.56);

\path[draw=drawColor,line width= 0.6pt,line join=round] (104.52, 23.26) --
	(104.52,105.56);

\path[draw=drawColor,line width= 0.6pt,line join=round] (133.03, 23.26) --
	(133.03,105.56);

\path[draw=drawColor,line width= 0.6pt,line join=round] (161.53, 23.26) --
	(161.53,105.56);

\path[draw=drawColor,line width= 0.6pt,line join=round] (190.04, 23.26) --
	(190.04,105.56);

\path[draw=drawColor,line width= 0.6pt,line join=round] (218.54, 23.26) --
	(218.54,105.56);
\definecolor{drawColor}{RGB}{51,34,136}
\definecolor{fillColor}{RGB}{51,34,136}

\path[draw=drawColor,line width= 0.4pt,line join=round,line cap=round,fill=fillColor] ( 47.51, 55.63) circle (  1.43);

\path[draw=drawColor,line width= 0.6pt,line join=round] ( 47.51, 45.86) -- ( 47.51, 52.66);

\path[draw=drawColor,line width= 0.6pt,line join=round] ( 47.51, 40.46) -- ( 47.51, 34.58);
\definecolor{fillColor}{RGB}{255,255,255}

\path[draw=drawColor,line width= 0.6pt,line join=round,line cap=round,fill=fillColor] ( 36.82, 45.86) --
	( 36.82, 40.46) --
	( 58.20, 40.46) --
	( 58.20, 45.86) --
	( 36.82, 45.86) --
	cycle;

\path[draw=drawColor,line width= 1.1pt,line join=round] ( 36.82, 43.15) -- ( 58.20, 43.15);

\path[draw=drawColor,line width= 0.6pt,line join=round] ( 76.02, 42.62) -- ( 76.02, 54.77);

\path[draw=drawColor,line width= 0.6pt,line join=round] ( 76.02, 32.90) -- ( 76.02, 27.14);

\path[draw=drawColor,line width= 0.6pt,line join=round,line cap=round,fill=fillColor] ( 65.33, 42.62) --
	( 65.33, 32.90) --
	( 86.71, 32.90) --
	( 86.71, 42.62) --
	( 65.33, 42.62) --
	cycle;

\path[draw=drawColor,line width= 1.1pt,line join=round] ( 65.33, 36.71) -- ( 86.71, 36.71);

\path[draw=drawColor,line width= 0.6pt,line join=round] (104.52, 44.34) -- (104.52, 50.15);

\path[draw=drawColor,line width= 0.6pt,line join=round] (104.52, 35.65) -- (104.52, 27.05);

\path[draw=drawColor,line width= 0.6pt,line join=round,line cap=round,fill=fillColor] ( 93.83, 44.34) --
	( 93.83, 35.65) --
	(115.21, 35.65) --
	(115.21, 44.34) --
	( 93.83, 44.34) --
	cycle;

\path[draw=drawColor,line width= 1.1pt,line join=round] ( 93.83, 38.01) -- (115.21, 38.01);
\definecolor{fillColor}{RGB}{51,34,136}

\path[draw=drawColor,line width= 0.4pt,line join=round,line cap=round,fill=fillColor] (133.03, 87.19) circle (  1.43);

\path[draw=drawColor,line width= 0.4pt,line join=round,line cap=round,fill=fillColor] (133.03, 86.74) circle (  1.43);

\path[draw=drawColor,line width= 0.4pt,line join=round,line cap=round,fill=fillColor] (133.03,101.82) circle (  1.43);

\path[draw=drawColor,line width= 0.4pt,line join=round,line cap=round,fill=fillColor] (133.03, 52.21) circle (  1.43);

\path[draw=drawColor,line width= 0.4pt,line join=round,line cap=round,fill=fillColor] (133.03, 53.88) circle (  1.43);

\path[draw=drawColor,line width= 0.4pt,line join=round,line cap=round,fill=fillColor] (133.03, 87.35) circle (  1.43);

\path[draw=drawColor,line width= 0.4pt,line join=round,line cap=round,fill=fillColor] (133.03, 59.49) circle (  1.43);

\path[draw=drawColor,line width= 0.4pt,line join=round,line cap=round,fill=fillColor] (133.03, 83.00) circle (  1.43);

\path[draw=drawColor,line width= 0.4pt,line join=round,line cap=round,fill=fillColor] (133.03, 84.96) circle (  1.43);

\path[draw=drawColor,line width= 0.4pt,line join=round,line cap=round,fill=fillColor] (133.03, 65.66) circle (  1.43);

\path[draw=drawColor,line width= 0.4pt,line join=round,line cap=round,fill=fillColor] (133.03, 97.59) circle (  1.43);

\path[draw=drawColor,line width= 0.6pt,line join=round] (133.03, 40.68) -- (133.03, 50.38);

\path[draw=drawColor,line width= 0.6pt,line join=round] (133.03, 33.96) -- (133.03, 27.00);
\definecolor{fillColor}{RGB}{255,255,255}

\path[draw=drawColor,line width= 0.6pt,line join=round,line cap=round,fill=fillColor] (122.34, 40.68) --
	(122.34, 33.96) --
	(143.72, 33.96) --
	(143.72, 40.68) --
	(122.34, 40.68) --
	cycle;

\path[draw=drawColor,line width= 1.1pt,line join=round] (122.34, 36.31) -- (143.72, 36.31);

\path[draw=drawColor,line width= 0.6pt,line join=round] (161.53, 39.65) -- (161.53, 48.48);

\path[draw=drawColor,line width= 0.6pt,line join=round] (161.53, 33.68) -- (161.53, 27.01);

\path[draw=drawColor,line width= 0.6pt,line join=round,line cap=round,fill=fillColor] (150.84, 39.65) --
	(150.84, 33.68) --
	(172.22, 33.68) --
	(172.22, 39.65) --
	(150.84, 39.65) --
	cycle;

\path[draw=drawColor,line width= 1.1pt,line join=round] (150.84, 37.95) -- (172.22, 37.95);

\path[draw=drawColor,line width= 0.6pt,line join=round] (190.04, 46.39) -- (190.04, 55.83);

\path[draw=drawColor,line width= 0.6pt,line join=round] (190.04, 35.98) -- (190.04, 27.12);

\path[draw=drawColor,line width= 0.6pt,line join=round,line cap=round,fill=fillColor] (179.35, 46.39) --
	(179.35, 35.98) --
	(200.73, 35.98) --
	(200.73, 46.39) --
	(179.35, 46.39) --
	cycle;

\path[draw=drawColor,line width= 1.1pt,line join=round] (179.35, 42.56) -- (200.73, 42.56);

\path[draw=drawColor,line width= 0.6pt,line join=round] (218.54, 46.33) -- (218.54, 51.52);

\path[draw=drawColor,line width= 0.6pt,line join=round] (218.54, 38.17) -- (218.54, 28.39);

\path[draw=drawColor,line width= 0.6pt,line join=round,line cap=round,fill=fillColor] (207.85, 46.33) --
	(207.85, 38.17) --
	(229.23, 38.17) --
	(229.23, 46.33) --
	(207.85, 46.33) --
	cycle;

\path[draw=drawColor,line width= 1.1pt,line join=round] (207.85, 44.57) -- (229.23, 44.57);
\definecolor{drawColor}{gray}{0.20}

\path[draw=drawColor,line width= 0.6pt,line join=round,line cap=round] ( 30.41, 23.26) rectangle (235.65,105.56);
\end{scope}
\begin{scope}
\path[clip] (  0.00,  0.00) rectangle (238.49,108.41);
\definecolor{drawColor}{gray}{0.30}

\node[text=drawColor,anchor=base east,inner sep=0pt, outer sep=0pt, scale=  0.64] at ( 25.46, 24.77) {0};

\node[text=drawColor,anchor=base east,inner sep=0pt, outer sep=0pt, scale=  0.64] at ( 25.46, 52.62) {500};

\node[text=drawColor,anchor=base east,inner sep=0pt, outer sep=0pt, scale=  0.64] at ( 25.46, 80.46) {1000};
\end{scope}
\begin{scope}
\path[clip] (  0.00,  0.00) rectangle (238.49,108.41);
\definecolor{drawColor}{gray}{0.20}

\path[draw=drawColor,line width= 0.6pt,line join=round] ( 27.66, 26.98) --
	( 30.41, 26.98);

\path[draw=drawColor,line width= 0.6pt,line join=round] ( 27.66, 54.82) --
	( 30.41, 54.82);

\path[draw=drawColor,line width= 0.6pt,line join=round] ( 27.66, 82.66) --
	( 30.41, 82.66);
\end{scope}
\begin{scope}
\path[clip] (  0.00,  0.00) rectangle (238.49,108.41);
\definecolor{drawColor}{gray}{0.20}

\path[draw=drawColor,line width= 0.6pt,line join=round] ( 47.51, 20.51) --
	( 47.51, 23.26);

\path[draw=drawColor,line width= 0.6pt,line join=round] ( 76.02, 20.51) --
	( 76.02, 23.26);

\path[draw=drawColor,line width= 0.6pt,line join=round] (104.52, 20.51) --
	(104.52, 23.26);

\path[draw=drawColor,line width= 0.6pt,line join=round] (133.03, 20.51) --
	(133.03, 23.26);

\path[draw=drawColor,line width= 0.6pt,line join=round] (161.53, 20.51) --
	(161.53, 23.26);

\path[draw=drawColor,line width= 0.6pt,line join=round] (190.04, 20.51) --
	(190.04, 23.26);

\path[draw=drawColor,line width= 0.6pt,line join=round] (218.54, 20.51) --
	(218.54, 23.26);
\end{scope}
\begin{scope}
\path[clip] (  0.00,  0.00) rectangle (238.49,108.41);
\definecolor{drawColor}{gray}{0.30}

\node[text=drawColor,anchor=base,inner sep=0pt, outer sep=0pt, scale=  0.64] at ( 47.51, 13.90) {AF};

\node[text=drawColor,anchor=base,inner sep=0pt, outer sep=0pt, scale=  0.64] at ( 76.02, 13.90) {AS};

\node[text=drawColor,anchor=base,inner sep=0pt, outer sep=0pt, scale=  0.64] at (104.52, 13.90) {CN};

\node[text=drawColor,anchor=base,inner sep=0pt, outer sep=0pt, scale=  0.64] at (133.03, 13.90) {EU};

\node[text=drawColor,anchor=base,inner sep=0pt, outer sep=0pt, scale=  0.64] at (161.53, 13.90) {NA};

\node[text=drawColor,anchor=base,inner sep=0pt, outer sep=0pt, scale=  0.64] at (190.04, 13.90) {OC};

\node[text=drawColor,anchor=base,inner sep=0pt, outer sep=0pt, scale=  0.64] at (218.54, 13.90) {SA};
\end{scope}
\begin{scope}
\path[clip] (  0.00,  0.00) rectangle (238.49,108.41);
\definecolor{drawColor}{RGB}{0,0,0}

\node[text=drawColor,anchor=base,inner sep=0pt, outer sep=0pt, scale=  0.80] at (133.03,  4.40) {Peer Region};
\end{scope}
\begin{scope}
\path[clip] (  0.00,  0.00) rectangle (238.49,108.41);
\definecolor{drawColor}{RGB}{0,0,0}

\node[text=drawColor,rotate= 90.00,anchor=base,inner sep=0pt, outer sep=0pt, scale=  0.80] at (  8.36, 64.41) {Avg. RTT (ms)};
\end{scope}
\end{tikzpicture}
    \caption{Latency distribution of Lightning's peer-to-peer network}
    \label{fig:plot_latencies}
\end{figure}
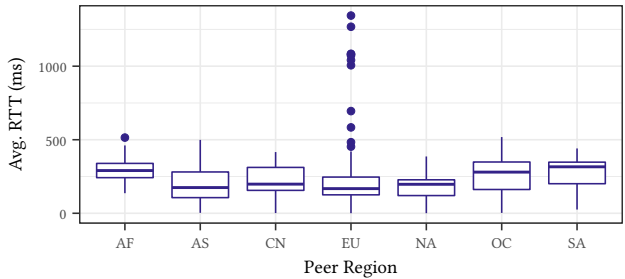

Based on this data, we setup a measurement study to infer a suitable latency
model for Lightning's peer-to-peer network. For this, we deployed seven
measurement nodes as close as possible to the aforementioned regional
clusters, \ie, in the following Amazon AWS regions: \texttt{us-west-1} (NA), \texttt{sa-east-1} (SA), \texttt{eu-central-1} (EU), \texttt{ap-southeast-2} (OC), \texttt{ap-south-1}
(AS), \texttt{me-south-1} (AF), and \texttt{ap-east-1} (CN). After
initialization, each measurement node starts collecting ICMP \texttt{ping} results to
each of the public Lightning IP addresses. In particular, each measurement
would send 100 \texttt{ping} requests to each Lightning node for 100 times,
which allows to build a more reliable round-trip time (RTT) model by averaging
over the results. Of the $2679$ addresses, we found $1297$ peers to be offline
or not reachable via ICMP, which corresponds to around $48\%$ of the network. 
The regional latency distribution for the remaining peers is shown in
Figure~\ref{fig:plot_latencies}: while there are some regional differences and
outliers,
the inter-peer latencies almost all fall below the 500ms mark, with the global
median being located around 250ms. 

\begin{figure}[]
    \centering
    \begin{tikzpicture}[x=1pt,y=1pt]
\definecolor{fillColor}{RGB}{255,255,255}
\path[use as bounding box,fill=fillColor,fill opacity=0.00] (0,0) rectangle (238.49,108.41);
\begin{scope}
\path[clip] (  0.00,  0.00) rectangle (238.49,108.41);
\definecolor{drawColor}{RGB}{255,255,255}
\definecolor{fillColor}{RGB}{255,255,255}

\path[draw=drawColor,line width= 0.6pt,line join=round,line cap=round,fill=fillColor] ( -0.00,  0.00) rectangle (238.49,108.41);
\end{scope}
\begin{scope}
\path[clip] ( 27.21, 23.26) rectangle (235.65,105.56);
\definecolor{fillColor}{RGB}{255,255,255}

\path[fill=fillColor] ( 27.21, 23.26) rectangle (235.65,105.56);
\definecolor{drawColor}{gray}{0.92}

\path[draw=drawColor,line width= 0.3pt,line join=round] ( 60.37, 23.26) --
	( 60.37,105.56);

\path[draw=drawColor,line width= 0.3pt,line join=round] (107.74, 23.26) --
	(107.74,105.56);

\path[draw=drawColor,line width= 0.3pt,line join=round] (155.11, 23.26) --
	(155.11,105.56);

\path[draw=drawColor,line width= 0.3pt,line join=round] (202.49, 23.26) --
	(202.49,105.56);

\path[draw=drawColor,line width= 0.6pt,line join=round] ( 27.21, 30.12) --
	(235.65, 30.12);

\path[draw=drawColor,line width= 0.6pt,line join=round] ( 27.21, 41.55) --
	(235.65, 41.55);

\path[draw=drawColor,line width= 0.6pt,line join=round] ( 27.21, 52.98) --
	(235.65, 52.98);

\path[draw=drawColor,line width= 0.6pt,line join=round] ( 27.21, 64.41) --
	(235.65, 64.41);

\path[draw=drawColor,line width= 0.6pt,line join=round] ( 27.21, 75.84) --
	(235.65, 75.84);

\path[draw=drawColor,line width= 0.6pt,line join=round] ( 27.21, 87.27) --
	(235.65, 87.27);

\path[draw=drawColor,line width= 0.6pt,line join=round] ( 27.21, 98.70) --
	(235.65, 98.70);

\path[draw=drawColor,line width= 0.6pt,line join=round] ( 36.68, 23.26) --
	( 36.68,105.56);

\path[draw=drawColor,line width= 0.6pt,line join=round] ( 84.05, 23.26) --
	( 84.05,105.56);

\path[draw=drawColor,line width= 0.6pt,line join=round] (131.43, 23.26) --
	(131.43,105.56);

\path[draw=drawColor,line width= 0.6pt,line join=round] (178.80, 23.26) --
	(178.80,105.56);

\path[draw=drawColor,line width= 0.6pt,line join=round] (226.17, 23.26) --
	(226.17,105.56);
\definecolor{fillColor}{RGB}{51,34,136}

\path[fill=fillColor] ( 36.68, 24.98) rectangle ( 37.77, 35.26);

\path[fill=fillColor] ( 36.68, 36.41) rectangle ( 38.64, 46.69);

\path[fill=fillColor] ( 36.68, 47.84) rectangle ( 39.22, 58.12);

\path[fill=fillColor] ( 36.68, 59.27) rectangle ( 40.02, 69.55);

\path[fill=fillColor] ( 36.68, 70.70) rectangle ( 45.68, 80.98);

\path[fill=fillColor] ( 36.68, 82.13) rectangle (115.13, 92.41);

\path[fill=fillColor] ( 36.68, 93.56) rectangle (129.79,103.85);
\definecolor{drawColor}{RGB}{0,0,0}

\node[text=drawColor,anchor=base west,inner sep=0pt, outer sep=0pt, scale=  0.85] at ( 42.27, 27.18) {0.6{\%}};

\node[text=drawColor,anchor=base west,inner sep=0pt, outer sep=0pt, scale=  0.85] at ( 43.15, 38.61) {1.0{\%}};

\node[text=drawColor,anchor=base west,inner sep=0pt, outer sep=0pt, scale=  0.85] at ( 43.73, 50.04) {1.3{\%}};

\node[text=drawColor,anchor=base west,inner sep=0pt, outer sep=0pt, scale=  0.85] at ( 44.52, 61.47) {1.8{\%}};

\node[text=drawColor,anchor=base west,inner sep=0pt, outer sep=0pt, scale=  0.85] at ( 50.19, 72.90) {4.7{\%}};

\node[text=drawColor,anchor=base west,inner sep=0pt, outer sep=0pt, scale=  0.85] at (120.70, 84.33) {41.4{\%}};

\node[text=drawColor,anchor=base west,inner sep=0pt, outer sep=0pt, scale=  0.85] at (135.36, 95.76) {49.1{\%}};
\definecolor{drawColor}{gray}{0.20}

\path[draw=drawColor,line width= 0.6pt,line join=round,line cap=round] ( 27.21, 23.26) rectangle (235.65,105.56);
\end{scope}
\begin{scope}
\path[clip] (  0.00,  0.00) rectangle (238.49,108.41);
\definecolor{drawColor}{gray}{0.30}

\node[text=drawColor,anchor=base east,inner sep=0pt, outer sep=0pt, scale=  0.64] at ( 22.26, 27.92) {AF};

\node[text=drawColor,anchor=base east,inner sep=0pt, outer sep=0pt, scale=  0.64] at ( 22.26, 39.35) {CN};

\node[text=drawColor,anchor=base east,inner sep=0pt, outer sep=0pt, scale=  0.64] at ( 22.26, 50.78) {SA};

\node[text=drawColor,anchor=base east,inner sep=0pt, outer sep=0pt, scale=  0.64] at ( 22.26, 62.21) {OC};

\node[text=drawColor,anchor=base east,inner sep=0pt, outer sep=0pt, scale=  0.64] at ( 22.26, 73.64) {AS};

\node[text=drawColor,anchor=base east,inner sep=0pt, outer sep=0pt, scale=  0.64] at ( 22.26, 85.07) {NA};

\node[text=drawColor,anchor=base east,inner sep=0pt, outer sep=0pt, scale=  0.64] at ( 22.26, 96.50) {EU};
\end{scope}
\begin{scope}
\path[clip] (  0.00,  0.00) rectangle (238.49,108.41);
\definecolor{drawColor}{gray}{0.20}

\path[draw=drawColor,line width= 0.6pt,line join=round] ( 24.46, 30.12) --
	( 27.21, 30.12);

\path[draw=drawColor,line width= 0.6pt,line join=round] ( 24.46, 41.55) --
	( 27.21, 41.55);

\path[draw=drawColor,line width= 0.6pt,line join=round] ( 24.46, 52.98) --
	( 27.21, 52.98);

\path[draw=drawColor,line width= 0.6pt,line join=round] ( 24.46, 64.41) --
	( 27.21, 64.41);

\path[draw=drawColor,line width= 0.6pt,line join=round] ( 24.46, 75.84) --
	( 27.21, 75.84);

\path[draw=drawColor,line width= 0.6pt,line join=round] ( 24.46, 87.27) --
	( 27.21, 87.27);

\path[draw=drawColor,line width= 0.6pt,line join=round] ( 24.46, 98.70) --
	( 27.21, 98.70);
\end{scope}
\begin{scope}
\path[clip] (  0.00,  0.00) rectangle (238.49,108.41);
\definecolor{drawColor}{gray}{0.20}

\path[draw=drawColor,line width= 0.6pt,line join=round] ( 36.68, 20.51) --
	( 36.68, 23.26);

\path[draw=drawColor,line width= 0.6pt,line join=round] ( 84.05, 20.51) --
	( 84.05, 23.26);

\path[draw=drawColor,line width= 0.6pt,line join=round] (131.43, 20.51) --
	(131.43, 23.26);

\path[draw=drawColor,line width= 0.6pt,line join=round] (178.80, 20.51) --
	(178.80, 23.26);

\path[draw=drawColor,line width= 0.6pt,line join=round] (226.17, 20.51) --
	(226.17, 23.26);
\end{scope}
\begin{scope}
\path[clip] (  0.00,  0.00) rectangle (238.49,108.41);
\definecolor{drawColor}{gray}{0.30}

\node[text=drawColor,anchor=base,inner sep=0pt, outer sep=0pt, scale=  0.64] at ( 36.68, 13.90) {0{\%}};

\node[text=drawColor,anchor=base,inner sep=0pt, outer sep=0pt, scale=  0.64] at ( 84.05, 13.90) {25{\%}};

\node[text=drawColor,anchor=base,inner sep=0pt, outer sep=0pt, scale=  0.64] at (131.43, 13.90) {50{\%}};

\node[text=drawColor,anchor=base,inner sep=0pt, outer sep=0pt, scale=  0.64] at (178.80, 13.90) {75{\%}};

\node[text=drawColor,anchor=base,inner sep=0pt, outer sep=0pt, scale=  0.64] at (226.17, 13.90) {100{\%}};
\end{scope}
\begin{scope}
\path[clip] (  0.00,  0.00) rectangle (238.49,108.41);
\definecolor{drawColor}{RGB}{0,0,0}

\node[text=drawColor,anchor=base,inner sep=0pt, outer sep=0pt, scale=  0.80] at (131.43,  4.40) {Share};
\end{scope}
\begin{scope}
\path[clip] (  0.00,  0.00) rectangle (238.49,108.41);
\definecolor{drawColor}{RGB}{0,0,0}

\node[text=drawColor,rotate= 90.00,anchor=base,inner sep=0pt, outer sep=0pt, scale=  0.80] at (  8.36, 64.41) {Region};
\end{scope}
\end{tikzpicture}
    \caption{Geographical peer distribution of the Lightning Network}
    \label{fig:plot_peers}
\end{figure}
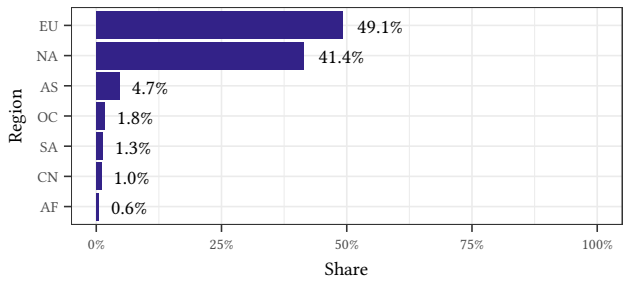

\end{document}